\documentclass[12pt]{article}

\usepackage{slashed}
\usepackage{amsmath}
\usepackage{amssymb}
\usepackage{lscape}
\newcommand{\minipagewidth}{160mm}
\newcommand{\ft}[2]{{\textstyle\frac{#1}{#2}}}

\csname @addtoreset\endcsname{equation}{section}

\newcommand{\SU}{SU}

\newcommand{\U}{U}
\newcommand{\USp}{USp}

\newcommand{\Gl}{G\ell }
\def\a{\alpha}

\def\e{\epsilon}

\def\p{\psi}

\def\m{\mu}
\def\n{\nu}

\newif\ifpdf
\ifx\pdfoutput\undefined
   \pdffalse
   \usepackage{cite}
  \usepackage{epsfig}
 \else
   \pdfoutput=1
   \pdftrue
  \usepackage[pdftex]{hyperref,graphicx}
\fi

\begin{document}

%%%%%%%%%%%%%%%%%%%%%%%%%%%%%%%%%%%%%%%%%%%%%%%%%%%%%%%%%%%
\begin{titlepage}
\begin{flushright}
UG-03-08\\
KUL-TF-04/06\\
SPIN-03/39\\
ITP-UU-03/58\\
hep-th/0403045
\end{flushright}
\vspace{.5cm}
\begin{center}
\baselineskip=16pt
{\LARGE    $N=2$ supergravity in five dimensions revisited  % \\ \vskip 0.2cm between lines
}\\
\vfill
{\large  Eric Bergshoeff$^1$, Sorin Cucu$^2$, Tim de Wit$^1$, \\[2mm]
Jos Gheerardyn$^2$, Stefan Vandoren$^3$ \\[2mm]
and Antoine Van Proeyen$^2$
 } \\
\vfill
{\small  $^1$ Center for Theoretical Physics, University of Groningen,\\
       Nijenborgh 4, 9747 AG Groningen, The Netherlands. \\
       \{e.a.bergshoeff, t.c.de.wit\}@phys.rug.nl \\[3mm]
  $^2$ Instituut voor Theoretische Fysica, Katholieke Universiteit Leuven,\\
       Celestijnenlaan 200D B-3001 Leuven, Belgium. \\
       sorin.cucu@ua.ac.be, \{jos.gheerardyn, antoine.vanproeyen\}@fys.kuleuven.ac.be \\[3mm]
  $^3$ Institute for Theoretical Physics, Utrecht University, \\
 Leuvenlaan 4, 3508 TA Utrecht, The Netherlands. \\
      s.vandoren@phys.uu.nl
 }
\end{center}
\vfill
\begin{center}
{\bf Abstract}
\end{center}
{\small We construct matter-coupled $N=2$ supergravity in five
dimensions, using the superconformal approach. For the matter sector we
take an arbitrary number of vector, tensor and hypermultiplets. By
allowing off-diagonal vector-tensor couplings we find more general
results than currently known in the literature. Our results provide the
appropriate starting point for a systematic search for BPS solutions, and
for applications of M-theory compactifications on Calabi-Yau manifolds
with fluxes. }
\end{titlepage}
\tableofcontents{}
\newpage

%%%%%%%%%%%%%%%%%%%%%%%%%%%%%%%%%%%%%%%%%%%%%%%%%%%%%%%%%%%%%%%%%%%%%%%%%%%%%%%
\section{Introduction\label{s:intr}}
%%%%%%%%%%%%%%%%%%%%%%%%%%%%%%%%%%%%%%%%%%%%%%%%%%%%%%%%%%%%%%%%%%%%%%%%%%%%%%%

Matter-coupled supergravities in five dimensions have attracted renewed
attention \cite{Gunaydin:1999zx,Ceresole:2000jd} due to the important
role they play in the Randall-Sundrum (RS) braneworld scenario
\cite{Randall:1999vf,Randall:1999ee}, in M-theory compactifications on
Calabi-Yau manifolds \cite{Cadavid:1995bk} with applications to flop
transitions and cosmology \cite{Jarv:2003qx,Jarv:2003qy},
 and in  the ${\rm AdS}_6/{\rm CFT}_5$
\cite{Nishimura:2000wj,D'Auria:2000ad} and ${\rm AdS}_5/{\rm CFT}_4$
\cite{Balasubramanian:2000pq} correspondences. Generically, one is
interested in studying BPS solutions, such as domain walls or black
holes, or in vacua obtained from flux-compactifications, perhaps even in
time-dependent solutions with a non-vanishing cosmological constant. In
many of these applications, a crucial role is played by the properties of
the scalar potential that appears after coupling matter multiplets to
$N=2$ supergravity. For example, the possibility for finding a
supersymmetric RS scenario depends on the existence of a domain-wall
solution (which requires a scalar potential) containing a warp factor
with the correct asymptotic behaviour such that gravity is suppressed in
the transverse direction. It is proved
in~\cite{Ceresole:2001wi,Alekseevsky:2001if} that one cannot restrict
oneself to vector multiplets, but that hypermultiplets are needed. Some
interesting solutions have already been found in
\cite{Behrndt:2001km,Anguelova:2002gd}.

In order to make a more systematic search for the variety of BPS
solutions such as domain walls or black holes that appear in five
dimensions, we need to know the most general form of matter-coupled
supergravity. It is the purpose of this work to construct these matter
couplings. Actually, the present work is the third and last in  a series
of papers where we apply the superconformal programme to derive these
matter couplings  (see also
\cite{Kugo:2000af,Kugo:2000hn,Fujita:2001kv}). The first paper dealt with
the construction of the $N=2$ conformal supergravity multiplet
\cite{Bergshoeff:2001hc},  while in the second paper we presented the
superconformal matter couplings \cite{Bergshoeff:2002qk}. In the present
work we will perform the last step in the superconformal programme,
i.e.~perform the gauge-fixing and obtain the matter-coupled $N=2$, $D=5$
Poincar{\'e} supergravity theory.

Quite some work on $N=2$ matter-coupled supergravities in five dimensions
has already been done. Since we claim to give more general results than
currently known in the literature, we first summarize what has been done.
The pure supergravity sector was constructed in \cite{Cremmer:1980gs}.
The coupling to vector multiplets was given in
\cite{Gunaydin:1984bi,Gunaydin:1985ak}. More recently, the addition of
tensor multiplets was considered in
\cite{Gunaydin:1999zx,Gunaydin:2000xk}. There it was stated that certain
couplings between vector and tensor multiplets were impossible to
supersymmetrize in a gauge-invariant way (except possibly in very special
cases). In this work we will show that such couplings are possible and
can be supersymmetrized thereby generalizing the results of
\cite{Gunaydin:1999zx,Gunaydin:2000xk}. Finally, vector, tensor and
hypermultiplets were treated together in \cite{Ceresole:2000jd}.

The superconformal programme, apart from leading us to general matter
couplings, has another bonus. It is well known that the scalars of the
matter sector can be viewed as the coordinates on a manifold. It  turns
out that there are interesting relations between the geometries of these
scalar manifolds before and after gauge-fixing the superconformal
symmetries. In \cite{deWit:1999fp,deWit:2001dj}, followed by
\cite{Bergshoeff:2002qk}, this was demonstrated for hypermultiplet
scalars.  The geometries before gauge-fixing are hypercomplex or
hyper-K{\"a}hler dependent on whether there exists an action or not (see
\cite{Bergshoeff:2002qk} for more details). After gauge-fixing, the
relevant geometries are quaternionic and quaternionic-K{\"a}hler,
respectively. Since the relations between these geometries are
interesting in themselves, and can be studied independently of the
present context, we are preparing a companion paper where we present the
details about all the geometries involved and their relations
\cite{Bergshoeff:2003yy}. Sometimes we will refer in this paper to
\cite{Bergshoeff:2003yy} for more details on the geometry.

The conformal programme has already been discussed extensively at several
occasions (see e.g.~\cite{VanProeyen:1983wk,Bergshoeff:1986mz}) including
our previous paper \cite{Bergshoeff:2002qk}. We refer the reader to these
reviews for more details. We will attempt to give a flavour of the
conformal approach by presenting a toy model. More explicitly, consider a
scalar-gravity model in four dimensions. We start with a conformally
invariant action for a scalar field $\phi$
\begin{equation}
\label{eq:toymodel} {\mathcal L} = \sqrt{|g|} \left[  \ft12
(\partial\phi)^2 + \ft{1}{12} R \phi^2 \right] ,
\end{equation}
which is invariant under the following local dilatations (with parameter
$\Lambda_D(x)$)
\begin{equation}
 \delta\phi = \Lambda_D \phi , \qquad \delta g_{\mu\nu} = -2\Lambda_D g_{\mu\nu} .
\end{equation}
This dilatation symmetry can be gauge fixed by choosing the gauge
\begin{equation}\label{gf}
\phi=\frac{\sqrt6}{\kappa}.
\end{equation}
\noindent This leads to the Poincar{\'e} action
\begin{equation}
\label{eq:EH} {\mathcal L} = \frac{1}{2\kappa^2} \sqrt{|g|} R .
\end{equation}
Therefore, the actions~(\ref{eq:toymodel}) and~(\ref{eq:EH}) are {\sl
gauge equivalent}. Alternatively, we could have chosen new coordinates
($g'_{\m\n}=g_{\m\n}\kappa ^2\phi^2$), such that the resulting action is
manifestly invariant under the dilatation symmetry. Although $\phi$ still
transforms under dilatations, the field does not appear in the action
anymore. The scalar $\phi$ has no physical degrees of freedom, and is
called a `compensating scalar'. Note that the scalar kinetic term has the
wrong sign; this is a generic feature of compensating scalars which we
will also encounter in the more complicated case of conformal
supergravity.

The same mechanism will be used in this paper to obtain five-dimensional
matter-coupled Poincar{\'e} supergravity.  To this end, the Poincar{\'e} algebra
is first extended to the local superconformal algebra $F^2(4)$. In our
first paper~\cite{Bergshoeff:2001hc} we constructed the minimal
representation of the superconformal algebra containing the graviton,
called the standard Weyl multiplet. This multiplet plays the role of
$g_{\mu\nu}$ in the above toy model. It turns out that in the case of
$N=2$, $D=5$ supergravity we need one hypermultiplet and one vector
multiplet as compensators. They play the role of the compensating scalar
$\phi$ in the above toy model. We thus have
\begin{equation}
\begin{array}{ccl}
g_{\mu\nu} &\rightarrow& {\mbox {standard\ Weyl\ multiplet}},\\
\phi &\rightarrow& {\mbox {1\ hyper-\ +\ 1\ vector-multiplet}} .
\end{array}
\end{equation}

\noindent On top of this we add an arbitrary number of $n_V$ vector,
$n_T$ tensor and $n_H$ hypermultiplets. We thus end up with $(n_V+1)$
vector, $n_T$ tensor  and $(n_H+1)$ hypermultiplets. As explained
in~\cite{Gunaydin:1999zx,Gunaydin:2000xk,Bergshoeff:2002qk}, the tensor
generically is part of a vector-tensor multiplet which is a hybrid form
of a vector and a tensor multiplet. The label of the vector-tensor
multiplet has $n_V+1$ vector-multiplet and $n_T$ tensor-multiplet
directions. Our starting point is therefore a number of vector, tensor
and hypermultiplets coupled to conformal supergravity:
\begin{equation}
\label{eq:superconf-action} {\mathcal L}_{\rm Total} = {\mathcal L}_{\rm
Vector-Tensor} + {\mathcal L}_{\rm Hyper} .
\end{equation}
These conformal couplings, which are the analogue of (\ref{eq:toymodel})
in the toy model, have been constructed in our second paper
\cite{Bergshoeff:2002qk}, and are repeated in
appendix~\ref{app:conf-matter} to keep the presentation self-contained.
In the main part of this paper we will discuss the final step, i.e.~the
gauge-fixing, where we get rid of all the superconformal symmetries that
are not part of the super-Poincar{\'e} algebra. This is the analogue of
(\ref{gf}), leading to a result similar to (\ref{eq:EH}). The main goal
of this paper is to derive the analogue of (\ref{eq:EH}). In the present
case we end up with $n_V$ vector, $n_T$ tensor and $n_H$ hypermultiplets
coupled to $N=2$, $D=5$ Poincar{\'e} supergravity. The final answer is given
in (\ref{final}). For the actual purpose of searching for supersymmetric
BPS solutions we only need the bosonic terms in the action. These have
been collected in (\ref{truncated}).

The organization of this paper is as follows. In
section~\ref{s:multiplets}, we give the field content of the standard
Weyl multiplet and the different matter multiplets: vector, tensor and
hyper. The corresponding supersymmetry and superconformal transformation
rules, together with the invariant actions are given in appendix
\ref{app:conf-matter}. Next, some details about the conformal geometry
for hypermultiplets, i.e.~the geometry before gauge-fixing, and its
relation to the geometry after gauge-fixing, are presented in
section~\ref{s:confgeo}. The gauge-fixing procedure is discussed in
section~\ref{s:gf}, accompanied by some well-known properties of very
special geometry, listed in appendix~\ref{app:veryspecial}. The resulting
action after gauge-fixing is given in section~\ref{s:results}. In
section~\ref{s:earlier}, we compare our results with the existing
literature. Finally, in section~\ref{s:simplified} we collect those terms
in the action  and supersymmetry rules that will be relevant to our
search for BPS solutions.

We use the same notation as in our previous two papers except that the
sign of the spacetime Ricci tensor and Ricci scalar has changed. For the
convenience of the reader we have collected the definitions of all
curvatures in appendix~\ref{app:notations}. Further details of the
notation can be found in the appendix of our first
paper~\cite{Bergshoeff:2001hc}.

%%%%%%%%%%%%%%%%%%%%%%%%%%%%%%%%%%%%%%%%%%%%%%%%%%%%%%%%%%%%%%%%%%%%%%%%%%
\section{Multiplets}\label{s:multiplets}
%%%%%%%%%%%%%%%%%%%%%%%%%%%%%%%%%%%%%%%%%%%%%%%%%%%%%%%%%%%%%%%%%%%%%%%%%%

The fields of the standard Weyl multiplet and their properties are listed
in table~\ref{tbl:fieldsWeyls}. The full details of this multiplet are
given in \cite{Bergshoeff:2001hc}. We use the following notation for
indices: $\mu (a)$ are curved (flat) world indices with $\mu, a = 0,
1,2,3,4$ and $i=1,2$ is an $\SU(2)$ index. All fermions are symplectic
Majorana spinors.
\begin{table}[htb]
\begin{center}
$  \begin{array}{||c|cccc||} \hline\hline
\mbox{Field} & \#  & \mbox{Gauge}&\SU(2)& w \\
\hline \rule[-1mm]{0mm}{6mm}& \multicolumn{4}{c||}{\mbox{Elementary gauge
fields}} \\
\rule[-1mm]{0mm}{6mm} e_\mu{}^a & 9 & P^a &1&-1\\
\rule[-1mm]{0mm}{6mm}
b_\mu   & 0 & D &1&0\\
\rule[-1mm]{0mm}{6mm}
V_\mu^{(ij)} &12  & \SU(2)\,&3&0\\
\hline \rule[-1mm]{0mm}{6mm} \psi _\mu ^i  & 24 & Q^i &2&-\frac12
\\ [1mm] \hline\hline \rule[-1mm]{0mm}{6mm}&
\multicolumn{4}{c||}{\mbox{Dependent gauge
fields}} \\
\rule[-1mm]{0mm}{6mm} \omega_\mu{}^{ab} & - & M^{[ab]} &1&0\\
\rule[-1mm]{0mm}{6mm}
f_\mu{}^a   & - & K^a &1&1\\
\hline \rule[-1mm]{0mm}{6mm} \phi _\mu ^i  & - & S^i &2&\frac12 \\
[1mm] \hline\hline \rule[-1mm]{0mm}{6mm} &
\multicolumn{4}{c||}{\mbox{Matter fields
}}\\
\rule[-1mm]{0mm}{6mm} T_{[ab]}  & 10 &  &1&1\\
\rule[-1mm]{0mm}{6mm} D & 1 & &1&2\\ \hline \rule[-1mm]{0mm}{6mm} \chi ^i
& 8 & &2&\frac32\\ [1mm] \hline\hline
\end{array}$
\caption{\it Fields of the 32 + 32 standard Weyl multiplet. The symbol
$\#$ indicates the off-shell degrees of freedom. The first block contains
the (bosonic and fermionic) gauge fields of the superconformal algebra.
The fields in the middle block are dependent gauge fields.  The extra
matter fields that appear in the standard Weyl multiplet are displayed in
the lower block. Note that we have suppressed the spinor index of both
the $Q$ and $S$ generators, of the corresponding gauge fields and of the
matter field $\chi^i$.
 }\label{tbl:fieldsWeyls}
\end{center}
\end{table}
In \cite{Bergshoeff:2002qk} we constructed vector-tensor multiplets and
hypermultiplets in the background of this Weyl multiplet. Vector-tensor
multiplets are a hybrid form of vector and tensor multiplets. The field
content and further properties of these multiplets are given in
table~\ref{tbl:multiplets}. Here we have introduced the following
indices: $I= 0, \cdots , n_V$ labels the adjoint representation of some
gauge group $G$, and $M = n_V + 1, \cdots  , n_V + n_T$ labels some
representation of $G$, possibly reducible, under which the tensors
transform (see below for more details). Finally, $X = 1, \cdots , 4(n_H
+1)$ and $(i,A)$ with $i=1,2$ and $A = 1, \cdots , 2(n_H +1)$ are,
respectively, the curved and flat indices of the hypermultiplet scalar
manifold\footnote{For comparison with our previous
paper~\cite{Bergshoeff:2002qk}: $
  n=n_V+1,\qquad  m=n_T, \qquad r=n_H+1.
$}.
Note that we have introduced $n_V+1$ vector multiplets and $n_H+1$
hypermultiplets to indicate that one of the vector multiplets and one of
the hypermultiplets serve as compensating multiplets.
\begin{table}[htb]
\begin{minipage}{\minipagewidth}%{\linewidth}
\renewcommand{\thefootnote}{\thempfootnote}
\begin{center}
\begin{tabular}{||c|ccc||}
\hline \rule[-1mm]{0mm}{6mm}
Field       & $\SU(2)$ & $w$ & \# d.o.f.  \\
\hline
 & \multicolumn{3}{c||}{off-shell vector multiplet} \\
\rule[-1mm]{0mm}{6mm}
$A_\m^I$      & 1 & 0             & $4 (n_V+1)$ \\
\rule[-1mm]{0mm}{6mm}
$Y^{ij I}$    & 3 & 2             & $3 (n_V+1)$ \\
\rule[-1mm]{0mm}{6mm}
$\sigma^I$        & 1 & 1             & $1 (n_V+1)$ \\
\hline \rule[-1mm]{0mm}{6mm}
$\p^{i I}$      & 2 & $3/2$         & $8 (n_V+1)$ \\[1mm]
\hline \hline
 & \multicolumn{3}{c||}{on-shell tensor multiplet} \\
\rule[-1mm]{0mm}{6mm}
$B_{\m\n}^M$      & 1 & 0         & $3 n_T$ \\
\rule[-1mm]{0mm}{6mm}
$Y^{ij M}$    & 3 & 2             & $0$ \\
\rule[-1mm]{0mm}{6mm}
$\sigma ^M$        & 1 & 1             & $1 n_T$ \\
\hline \rule[-1mm]{0mm}{6mm}
$\psi ^{i M}$      & 2 & $3/2$         & $4 n_T$ \\[1mm]
\hline \hline
 & \multicolumn{3}{c||}{on-shell hypermultiplet
 }   \\
\rule[-1mm]{0mm}{6mm}
$q^X$      & 2 & $3/2$        & $4 (n_H+1)$ \\
\hline \rule[-1mm]{0mm}{6mm}
$\zeta^A$      & 1 & 2       & $4 (n_H+1)$ \\[1mm]
\hline
\end{tabular}
\caption{\it The relevant $D=5$ superconformal matter multiplets. We
introduce $(n_V+1)$ off-shell vector multiplets, $n_T$ on-shell tensor
multiplets and $(n_H+1)$ on-shell hypermultiplets. Indicated are their
$\SU(2)$ representations, Weyl weights $w$ and the number of
off-shell/on-shell degrees of freedom. Each of the multiplets describes 4
+ 4 on-shell degrees of freedom. } \label{tbl:multiplets}
\end{center}
\end{minipage}
\end{table}
When combining vectors and tensors into a vector-tensor multiplet we will
sometimes write $\tilde I = (I,M)$.

We first consider the vector-tensor multiplet. A remarkable feature of
the vector-tensor multiplet is that the vector-part is off-shell whereas
the tensor-part is on-shell. The gauge transformations of the
vector-tensor multiplet are specified by matrices  $(t_I){}_{\widetilde
J}{}^{\widetilde K}$ that satisfy the commutation relations (with
structure constants $f_{IJ}{}^K$)
\begin{equation}
[t_I, t_J] = - f_{IJ}{}^K t_K.
\end{equation}
These gauge transformations (with parameters $\Lambda^I$) are given by
\begin{equation}
\delta_G(\Lambda^J)A_\mu^I = \partial_\mu\Lambda^I +
gA_\mu^Jf_{JK}{}^I\Lambda^K, \qquad \delta_G(\Lambda^J)X^{\widetilde I} =
-g\Lambda^J (t_J)_{\widetilde K}{}^{\widetilde I}X^{\widetilde K},
\end{equation}
where $X^{\widetilde I}$ is a general matter field and $g$ is the
coupling constant of the group $G$. Closure of the supersymmetry algebra
requires $(t_J)_M{}^I=0$, so the generic form of the matrices $t_I$ is
given by
\begin{equation}
(t_I){}_{\widetilde J}{}^{\widetilde K} =
\begin{pmatrix}
f_{IJ}{}^K & (t_I)_J{}^N \\
0 & (t_I)_M{}^N
\end{pmatrix}, \qquad
\left\{
\begin{array}{ccl}
I,J,K &=& 0, \ldots ,n_V \\
M,N &=& n_V+1, \ldots ,n_V+n_T.
\end{array}
\right.  \label{matricest}
\end{equation}
Sometimes we extend the index $I$ in $(t_I)_{\widetilde J}{}^{\widetilde
K}$ to $\tilde I$ with the understanding that
\begin{equation}
(t_M){}_{\widetilde J}{}^{\widetilde K}  = 0.
\end{equation}
If $n_T \ne 0$ then the representation $(t_I){}_{\widetilde
J}{}^{\widetilde K}$ is reducible. In our second paper
\cite{Bergshoeff:2002qk} we showed that this representation can be more
general than assumed so far in treatments of vector-tensor couplings. In
particular, the off-diagonal matrix elements $(t_I)_J{}^N$  lead to new
matter couplings, and the requirement that $n_T$ is even will only appear
when we demand the existence of an action or if we require the absence of
tachyonic modes. The supersymmetry rules of the vector-tensor multiplet
can be found in appendix \ref{app:conf-matter}.

In the absence of an action, the vector-tensor multiplet is characterized
by the matrices $t_I$. In order to write down a superconformal action we
need to introduce two more symbols: a fully symmetric tensor
$C_{\widetilde I\widetilde J\widetilde K}$ and an antisymmetric and
invertible tensor $\Omega_{MN}$. They are related by
\begin{eqnarray}
   &   & C_{M\widetilde J \widetilde K} = t_{(\widetilde{J} \widetilde{K})}{}^P
\Omega_{ PM}, \nonumber\\
   &   &   t_{I[M}{}^P\Omega _{N]P}=0, \nonumber\\
   && t_{I(\widetilde J}{}^{\widetilde M} C_{\widetilde K\widetilde
L)\widetilde M} = 0 ,
 \label{COmegarelations}
\end{eqnarray}
see section~3.2 of \cite{Bergshoeff:2002qk}. The corresponding action is
given in appendix \ref{app:conf-matter}.

We now turn to the hypermultiplet. In the absence of an action, the
superconformal tensor calculus performed in~\cite{Bergshoeff:2002qk}
resulted in the construction of a hypercomplex manifold spanned by the
$(4n_H+4)$ hyperscalars $q^{X}$. This manifold includes the four scalars
of the compensating hypermultiplet. The geometrical properties of the
hypercomplex manifold are determined by a collection of vielbeins $
f_{X}^{iA}$, satisfying the following constraints:
\begin{eqnarray}\label{covconst}
 f^{i A}_{Y}  f^{ X}_{i A}
&=& \delta_{Y}^{ X} ,\qquad  f^{iA}_{ X}  f^{ X}_{j B} = \delta^i_j \delta^{ A}_{ B},\nonumber\\
{\mathfrak D}_{ Y}  f_{i B}^{X} &\equiv& \partial_{Y}  f_{i B}^{X} -
 \omega_{ Y B}^{\quad  A}  f_{iA}^{ X} +
 \Gamma_{ Z Y}^{\quad  X}  f_{i B}^{ Z} =
0,
\end{eqnarray}
where $ \Gamma_{ ZY}{}^{X}$ can be interpreted as the affine connection
on the manifold, and $ \omega_{ Y B}{}^{A}$ as the
$\Gl(n_H+1,\mathbb{H})$ connection. The last constraint shows that the
vielbeins are covariantly constant with respect to the connections
$\Gamma$ and $\omega$. These connections do not represent independent
functions, see below.  On this manifold we introduce a triplet of complex
structures, defined as
\begin{equation}
 {\vec J}_{ X}{}^{ Y} \equiv - {\rm i}  f_{ X}^{i A}\vec \sigma_i{}^j  f_{j A}^{ Y}.
\label{defJf}
\end{equation}
Using~(\ref{covconst}), they are covariantly constant and satisfy the
quaternion algebra, which is that for any vectors $\vec A$ and $\vec B$,
\begin{equation}
\vec A\cdot \vec J_X{}^Z\, \vec B\cdot \vec J_Z{}^Y = -\delta _X{}^Y\vec
A\cdot \vec B + (\vec A\times \vec B)\cdot \vec J_X{}^Y
  , \label{defJ}
\end{equation}
At some places we will use a doublet notation instead of the triplet (or
vector) notation:
\begin{equation}
 J_{X}{}^{ Y}{}_i{}^j \equiv {\rm i} {\vec J}{}_{ X}{}^{ Y} \cdot
{\vec \sigma}_i{}^j
 = 2  f^{j A}_{ X} f^{ Y}_{i A}-
\delta_i^j \delta_{ X}^{ Y}. \label{doublet-J}
\end{equation}
The same transition between doublet and triplet notation is also used for
other quantities in the adjoint representation of $\SU(2)$.

Note that the complex structure is obtained from the vielbeins. Its
covariant constancy is sufficient to determine the affine connection. The
latter is then called the Obata connection (similar to the definition of
a Levi-Civita connection determined from the covariant constancy of a
metric). Further, once this Obata connection is known, the covariant
constancy of the vielbeins as in the last line of~(\ref{covconst})
determines the $\Gl(n_H+1,\mathbb{H})$ connection $ \omega_{XA}{}^B$. In
this way all quantities can be derived from the vielbeins.

In order to build local supergravity theories, we require the hyperscalar
manifolds that we will use to be `conformal' in the sense that they
contain a homothetic Killing vector $k^X$, describing the dilatations of
$q^X$ \cite{deWit:1998zg,deWit:1999fp}
\begin{equation}
  {\mathfrak D}_Y k^X=\ft32\delta _Y^X.
\label{homothety}
\end{equation}
Using the complex structures this defines $\SU(2)$ symmetry generators $
{\vec k}^{ X}$:
\begin{equation}
\vec k^X\equiv \ft13 k^Y\vec
  J_Y{}^X.
 \label{kveck}
\end{equation}
Both $k^X$ and $ {\vec k}^{ X}$ are discussed in section~2.3.2
of~\cite{Bergshoeff:2002qk}.

In section~2.3.3 of~\cite{Bergshoeff:2002qk}, we also considered the
action of the symmetry group gauged by the vector multiplets on the
hypercomplex manifold. Their generators are therefore labelled by the
index $I$. They are parametrized by triholomorphic (i.e. leaving the
complex structures invariant) Killing\footnote{The word `Killing vector'
is in fact only appropriate when they respect a metric,
see~(\ref{Killing}), which is the case that we will mostly consider in
this paper.} vectors $k^X_I$:
\begin{eqnarray}
  \delta_G q^X&=& -g \Lambda_G^I k_I^X(q), \nonumber\\
 \delta_G \zeta^A &=& -g \Lambda^I_G t_{IB}{}^A(q) \zeta^B-\zeta ^B\omega _{XB}{}^A\delta_G q^X
 , \nonumber\\
&&t_{IB}{}^A\equiv \ft 12f^Y_{iB}\mathfrak{D}_Yk^X_If^{iA}_X, \qquad
f^{Y(i}_Af^{j)B}_X \mathfrak{D}_Y k^X_I=0. \label{isometries}
\end{eqnarray}
The supersymmetry rules of the hypermultiplet are given in
appendix~\ref{app:conf-matter}.

So far, we have not assumed the existence of an action. In order to write
down an action we need to introduce a covariantly constant antisymmetric
invertible tensor $C_{AB}$ which will be used to raise and lower the $A$
indices. We can now construct the metric on the scalar manifold as
\begin{equation}
g_{XY} = f_X^{iA} C_{AB} \epsilon_{ij} f_Y^{jB},
\end{equation}
and the hypercomplex manifold becomes a hyper-K{\"a}hler manifold. The Obata
connection then coincides with the Levi-Civita connection. The action
also contains four-Fermi terms, proportional to a tensor $W_{ABC}{}^D$.
The latter is defined in terms of the Riemann tensor and conversely
completely determines the Riemann tensor:
\begin{equation}\label{defstrW}
W_{ABC}{}^D = \ft12 f_A^{iX} f_{iB}^Y f_{jC}^Z f_W^{Dj} R_{XYZ}{}^W,
\qquad  R_{XYW}{}^Z = - \ft 12 f^{Ai}_X
\varepsilon_{ij}f^{jB}_Yf_W^{kC}f^Z_{kD} W_{ABC}{}^D,
\end{equation}
see appendix~B of \cite{Bergshoeff:2002qk}. The metric breaks the
holonomy group from $\Gl(n_H+1,\mathbb{H})$ to $\USp(2,2n_H)$, where we
chose the metric to have signature $(----++\ldots +)$, as it is required
for a physical theory with positive kinetic terms. The $-$ signs
correspond to the scalars of the compensating multiplet. This is to be
compared with the negative kinetic energy of the compensator in the toy
model discussed in the introduction. From the integrability condition
that follows from the covariant constancy of $C_{AB}$, and using a basis
with constant $C_{AB}$, we can determine that the connection $\omega_X$
is symmetric, and thus
\begin{equation}
  \omega _{XAB}\equiv \omega _{XA}{}^C C_{CB}=\omega _{XBA},
 \label{omegaSymp}
\end{equation}
is now the connection for $\USp(2,2n_H)$.

Finally, when an action exists, the symmetries should respect the metric,
i.e. the Killing equation
\begin{equation}
 {\mathfrak D}_{(X} k_{Y)I}=0 ,
 \label{Killing}
\end{equation}
should be satisfied. Then, moment maps $\vec P_I$ can be defined, see
section~3.3.2 of \cite{Bergshoeff:2002qk},
\begin{equation}
\partial_X\vec{P}_I=-\ft12 \vec{J}_{XY}k^Y_I,
\end{equation}
which by the conformal symmetry are determined to be
\begin{equation}\label{HK-moments}
\vec {P}_I = -\ft16 k^X {\vec J}_{X}{}^Y k^Z_I g_{YZ}.
\end{equation}
They appear in the scalar potential of the action, which is given in
appendix~\ref{app:conf-matter}. This discussion of isometries and moment
maps on such hyper-K{\"a}hler spaces applies as well in 6 and 4 spacetime
dimensions, and appeared in the 4-dimensional theories
in~\cite{Wit:2001bk}.
%%%%%%%%%%%%%%%%%%%%%%%%%%%%%%%%%%%%%%%%%%%%%%%%%%%%%%%%%%%%%%%%%%%%%%%%%%%%%%%
\section{Geometry\label{s:confgeo}}
%%%%%%%%%%%%%%%%%%%%%%%%%%%%%%%%%%%%%%%%%%%%%%%%%%%%%%%%%%%%%%%%%%%%%%%%%%%%%%%

In preparation for the later developments, but also as an introduction to
the geometrical aspects of the gauge-fixing procedure discussed
hereafter, we recall in this section some ideas concerning the connection
between conformal hypercomplex (hyper-K{\"a}hler) and quaternionic(-K{\"a}hler)
manifolds. The details of the explicit map between the corresponding
geometries are presented in the companion paper \cite{Bergshoeff:2003yy}.
\textit{For the rest of this paper we will always assume the presence of
a metric. Therefore, we will only deal with the hyper-K{\"a}hler and
quaternionic-K{\"a}hler geometries.} The geometry and relation between these
spaces were also analysed in \cite{deWit:2001dj}, and in the mathematics
literature~\cite{Swann}. Our analysis here involves a different choice of
coordinates and gauge-fixing procedure than in \cite{deWit:2001dj}, and
will turn out to be more convenient for our purposes. The case without a
metric, i.e.~the map between the hypercomplex and quaternionic
geometries, is discussed in \cite{Bergshoeff:2003yy}.

In the superconformal tensor calculus there is one vector and one
hypermultiplet that play a special role as a compensating multiplets. We
choose 3 scalars of the hypermultiplet to  gauge-fix the  three $\SU(2)$
gauge transformations and 1 to gauge-fix the dilatation $D$. This means
that 4 compensating scalars will be removed from the hyper-K{\"a}hler
manifold. In view of this, it is convenient to split these coordinates
off in a manifest way by making a specific coordinate choice on the
hyper-K{\"a}hler manifold. Full proofs and the exact mapping between
arbitrary hypercomplex and quaternionic spaces in this way will be
published in~\cite{Bergshoeff:2003yy}. Here we will summarize the
relevant results. From now on we will use the hat-notation for objects
that are defined on the `higher-dimensional' hyper-K{\"a}hler manifold. For
instance, the coordinates of the hyper-K{\"a}hler manifold are denoted by
$q^{\hat X}\ ({\hat X} = 1, \cdots ,4n_H+4)$.

The three isometries generated by the three $\SU(2)$ Killing vectors
$\hat {\vec k}{}^{\hat X}$ are gauged using the vectors of the Weyl
multiplet. Using Frobenius' theorem it is shown
in~\cite{Bergshoeff:2003yy} that the three-dimensional subspace spanned
by the directions of the three $\SU(2)$ transformations can be
parametrized by coordinates $z^\a$ (with $\alpha =1,2,3$). Furthermore,
using the homothetic Killing equation, one coordinate $z^0$ can be
associated with the dilatation transformation. The remaining directions
are indicated by $q^X$ ($X=1,\ldots,4n_H$). Thus, we split the
coordinates on the hyper-K{\"a}hler manifold as
\begin{equation}
\{q^{\hat X} \} = \{ z^0,\, z^\a,\, q^X \}.
\end{equation}
Throughout this paper we will work in this coordinate basis. In this
basis, the dilatation and $\SU(2)$ transformations, see~(\ref{homothety})
and~(\ref{kveck}), take on the following form~\cite{Bergshoeff:2003yy}:
\begin{equation}
  \hat k^{\hat X}(z,q) = \{ 3 z^0, 0, 0 \}, \qquad
  \hat {\vec k}^{\hat X}(z,q) = \{ 0, \vec k^\alpha(z,q), 0 \}.
\label{Killk}
\end{equation}
Thus in the bosonic sector only $z^0$ transforms under dilatations and
only $z^\alpha $ transforms under the $\SU(2)$ in the superconformal
group.

Similarly, we can split the tangent space index ${\hat A}$ as $\{\hat A
\} = \{ i, A\}$ ($i=1,2 ;\ A=1,\ldots,2n_H$), where $i$ is an $\SU(2)$
index, implying
\begin{equation}
  \{\zeta ^{\hat{A}}\} =\{ \zeta ^i,\,\zeta ^A\}.
 \label{fermionshypersplit}
\end{equation}
This basis is chosen such that in the fermionic sector only $\zeta ^i$
transforms under $S$-super\-symmetry.

Further analysis of the dilatations and SU(2) isometries shows that, in
these coordinates, the metric takes the form of a cone over a
tri-Sasakian manifold, see e.g.~\cite{Boyer:1998sf},
\begin{eqnarray}
{\rm d} \hat s^2
&\equiv& \hat g_{\hat X\hat Y} {\rm d} q^{\hat X} {\rm d} q^{\hat Y} , \nonumber\\
&=& -\frac{({\rm d} z^0)^2}{z^0} + z^0 \left\{ h_{XY} (q)\,{\rm d} q^X\,{\rm d} q^Y\right. \nonumber\\
&& \left.- g_{\alpha\beta} (z,q)\left[{\rm d} z^\alpha +
A_X^\alpha(z,q)\,{\rm d} q^X\right]\, \left[{\rm d} z^\beta + A_Y^\beta
(z,q)\,{\rm d} q^Y\right]\right\}  , \label{QK}
\end{eqnarray}
where we have chosen the signs and factors for later convenience. We have
defined
\begin{equation}
  A_X^\alpha(z,q)\equiv \hat f^\alpha_{ij} \hat f_X^{ij}=-\hat f^\alpha_{iA} \hat f_X^{iA}
,\qquad {\hat g}_{\alpha\beta}(z,q)\equiv z^0 g_{\alpha\beta}.
 \label{defAghat}
\end{equation}
The latter is an (invertible) metric in the $z^\alpha$-space, used to
raise and lower $\alpha,\beta$ indices. It turns out that, for each value
of $z^0$,
\begin{equation}
  g_{XY}(z,q)\equiv z^0h_{XY}(q)=\hat{g}_{XY}+ \hat{g}_{\alpha \beta
  }A_X^\alpha A_Y^\beta
 \label{defgXY}
\end{equation}
defines a quaternionic-K{\"a}hler metric on the base-space spanned by the
coordinates $q^X$.

The $\vec k^\alpha $ generate an $\SU(2)$ algebra, which is the statement
that
\begin{equation}
  \vec k^\gamma \times\partial _\gamma \vec k^\alpha= \vec k^\alpha.
 \label{SU2veck}
\end{equation}
${\vec k}_\alpha \equiv  {\hat g}_{\alpha\beta} {\vec k}^\beta$ are
proportional to the inverse of $\vec k^\alpha $ as $3\times 3$ matrices:
\begin{equation}
{\vec k}^\alpha \cdot {\vec k}_\beta = - z^0 \delta^\alpha_\beta.
\end{equation}
The dependence of these $\SU(2)$ Killing vectors and of $A_X^\alpha $ on
$z^0$ and $z^\alpha $ and $q^X$ is further restricted by
\begin{eqnarray}
   &   & \partial _0\vec k^\alpha =0,\qquad \partial _0\left( \frac{\vec k_\alpha }{z^0}
   \right) =0, \qquad \partial _0 \vec A_X =0, \qquad  \vec A_X\equiv \frac{1}{z^0}A_X^\alpha \vec
k_\alpha , \nonumber\\
   &   &   \left( z^0\partial _\alpha -\vec k_\alpha \times \right)
 \vec A_X=\partial_X\vec{k}_\alpha, \qquad
\partial _{[X}\vec A_{Y]}-\ft12\vec A_X\times \vec A_Y=\ft12\vec
J_{[X}{}^Zh_{Y]Z}.\label{diffeqn}
\end{eqnarray}

In this basis\footnote{We consider here the domain $z^0>0$ , which is
necessary to obtain at the end positive kinetic terms for the graviton.}
we find the following expressions for the vielbeins ${\hat f}_{\hat
X}^{i{\hat A}}$:
\begin{equation}
\begin{array}{lll}
 \hat f_0^{ij}={\rm i}\varepsilon^{ij}\sqrt{\frac{1}{2z^0}},\qquad & \hat f_\alpha^{ij}
  =  \sqrt{\frac{1}{2z^0}} \vec k_\alpha\cdot
\vec\sigma^{ij}
 ,\qquad &\hat f_X^{ij}= \sqrt{\frac{z^0}{2}} \vec A_X\cdot
\vec\sigma^{ij} , \\ [1mm] \hat f_0^{iA}=0,\qquad  & \hat
f_\alpha^{iA}=0,\qquad &\hat f_X^{iA}=f_X^{iA},
\end{array} \label{allf}
\end{equation}
where $f_X^{iA}(q)$ are the quaternionic-K{\"a}hler vielbeins. The inverse
vielbeins ${\hat f}^{\hat X}_{i{\hat A}}$ are given by
\begin{equation}
\begin{array}{lll}
\hat f^0_{ij} = -{\rm i}\varepsilon _{ij}\sqrt{\ft12z^0}, \qquad &
  \hat f^\alpha_{ij} =\sqrt{\frac{1}{2z^0}}\vec k^\alpha \cdot \vec
\sigma_{ij},\qquad &
 \hat f^X_{ij}=0,\\ [1mm]
\hat f^0_{iA}=0,\qquad & \hat{f}^\alpha_{iA}=-f^X_{iA}A_X^\alpha,\qquad &
\hat f_{iA}^X= f_{iA}^X,
\end{array}
\end{equation}
where $f^X_{iA}(q)$ are the inverse quaternionic-K{\"a}hler vielbeins. Note
that we have chosen our coordinates such that ${\hat f}^X_{ij} = 0$. This
means that from the supersymmetry rule $\delta q^{\hat X} = -{\rm i}
{\bar\epsilon}^i\zeta^{\hat A} f_{i\hat A}^{\hat X}$ it follows that
$q^X$ only transforms to $\zeta^A$ and not to $\zeta^i$.

For the complex structures we have
\begin{equation}
   \begin{array}{lll}
    \widehat{\vec J}_0{}^0=0 ,\qquad \qquad & \widehat{\vec J}_\alpha{}^0
    =\vec k_\alpha  , & \widehat{\vec J}_X{}^0=z^0\vec A_X  ,\\ [1mm]
    \widehat{\vec J}_0{}^\beta=\frac{1}{z^0}\vec k^\beta   ,& \widehat{\vec J}_\alpha{}^\beta=\frac{1}{z^0}\vec k_\alpha \times \vec k^\beta  ,\qquad&
    \widehat{\vec J}_X{}^\beta=\vec A_X \times \vec k^\beta
                               +\vec J_X{}^Z(\vec A_Z\cdot \vec k^\beta)  ,    \\ [1mm]
    \widehat{\vec J}_0{}^Y=0 ,& \widehat{\vec J}_\alpha{}^Y=0 ,& \widehat{\vec J}_X{}^Y=\vec
    J_X{}^Y.
  \end{array}
 \label{allhatJ}
\end{equation}
Finally, for the $\USp(2,2n_H)$ connections we find the following nonzero
components:
\begin{equation}
   \begin{array}{ll}
      & \hat{\omega }_{0 A}{}^B=\ft12f_Y^{iB}\partial_0 f_{iA}^Y+ \ft{1}{2z^0} \delta_A^B, \\ [1mm]
    \hat{\omega }_{\alpha i}{}^j=-{\rm i}\frac{1}{2z^0}\vec
k_\alpha \cdot\vec \sigma_i{}^j,\qquad   & \hat{\omega }_{\alpha A}{}^B=
\ft12 f_Y^{iB}\partial_\alpha f_{iA}^Y, \\ [1mm] \hat{\omega
}_{Xi}{}^j=A_X^\alpha \hat{\omega }_{\alpha i}{}^j= \omega_{Xi}{}^j, &
\hat{\omega }_{XA}{}^B = {\omega }_{XA}{}^B,\\ [1mm] \hat\omega_X{}_i{}^A
= {\rm i}\sqrt{\frac{1}{2z^0}} \varepsilon_{ik} f_X^{kA}, &
\hat{\omega}_{XA}{}^i= - {\rm i}\sqrt{\frac{z^0 }{2}} \varepsilon
^{ij}f_{jA}^Y h_{YX},
  \end{array}
 \label{allOmega}
\end{equation}
where ${\omega }_{XA}{}^B$ is defined in terms of the quaternionic-K{\"a}hler
vielbeins $f_X^{iA}$, the Levi-Civita connection computed with $z^0
h_{XY}$ and the $\SU(2)$ connection $\omega_{Xi}{}^j$ on the
quaternionic-K{\"a}hler manifold through the covariant constancy of the
vielbeins.\footnote{Note the subtle difference in notation: in the
hyper-K{\"a}hler manifold there is no $\SU(2)$ connection, and $\hat{\omega
}_{Xi}{}^j$ are components of the connection $\hat{\omega
}_{X\hat{A}}{}^{\hat{B}}$. On the other hand, in the quaternionic space
these components do not exist and $\omega_{Xi}{}^j={\rm i}\vec \omega
_X\cdot \vec \sigma_i{}^j$ is the $\SU(2)$ connection.} Using the vector
notation the $\SU(2)$ connection is given by
\begin{equation}
\vec\omega_X = -\ft{1}{2} \vec A_X .
\end{equation}

Using the above results, one can express any hatted quantity (geometric
quantity of the hyper-K{\"a}hler space) in terms of the unhatted ones
(geometric quantity of the quaternionic-K{\"a}hler space) and vice versa.
Here we list some explicit expressions for hatted quantities that occur
in the construction of the action,
\begin{eqnarray}
&& \hat C_{AB} =  C_{AB} , \qquad \hat C_{ij} = \varepsilon_{ij}, \qquad
\hat
C_{iA} = 0 , \nonumber\\
 && \widehat{\vec{P}}_I= \vec{P}_I, \nonumber\\
 && \hat k_I^{\hat X} = \{
0, -2 \vec k^\a \cdot (\vec\omega_X k_I^X - \ft{1}{z^0} \vec P_I) , k_I^X
\},
\nonumber\\
 && \widehat W_{ABC}{}^D = {\mathcal W}_{ABC}{}^D .
 \label{maphattedunhatted}
\end{eqnarray}
Here, $\vec{P}_I$ are the moment maps of the quaternionic-K{\"a}hler
manifold, in principle to be distinguished from the hyper-K{\"a}hler moment
maps defined in~(\ref{HK-moments}) (all hyper-K{\"a}hler relations from that
section should here be interpreted with hatted quantities and indices).
For $n_H\neq 0$ they are defined in terms of the Killing vectors and
complex structures as
\begin{equation}
  4n_H \vec P_I= z_0\vec J_X{}^Y{\mathfrak D}_Y k_I^X.
 \label{momentmap}
\end{equation}
In the absence of any physical hypermultiplets, i.e. $n_H=0$, moment maps
can still be present. In fact, they are constants, `Fayet-Iliopoulos (FI)
terms', restricted by the `equivariance condition'
\begin{equation}
  2\nu \vec P_I\times \vec P_J + f_{IJ}{}^K \vec P_K=0,
 \label{equivariance}
\end{equation}
where $\nu $ is not determined (multiplying again with $\nu$, this
becomes an equation for $\nu \vec P_I$). This will be further analysed in
section~\ref{s:results}. The moment maps~(\ref{momentmap}) satisfy a
similar equivariance condition~\cite{D'Auria:1991fj}.
Equations~(\ref{maphattedunhatted}) say that the hyper-K{\"a}hler moment maps
are the same as these moment maps, both when $n_H\neq 0$ and when
$n_H=0$. In the latter case, the FI terms are therefore related to the
gauge transformation of the scalars of the compensating hypermultiplet,
as is the case in 4 dimensions~\cite{deWit:1984pk}.

Similar to the $W_{ABCD}$ tensor (\ref{defstrW}) on a hyper-K{\"a}hler
manifold, we can introduce ${\cal W}_{ABCD}$, which is a completely
symmetric and traceless tensor. It is defined by the quaternionic-K{\"a}hler
Riemann tensor $R_{XYZ}{}^W$, and conversely it determines the latter, as
follows:
\begin{eqnarray}
{\cal W}_{AB}{}^{CD} &=& \ft12 L^{XY}{}_{BA}L_{WZ}{}^{CD}
 R_{XY}{}^{ZW}+2\frac{1}{z^0}\delta_{(A}{}^C\delta_{B)}{}^D, \qquad
 L^{XY}{}_{BA}\equiv f_A^{iX} f_{iB}^Y,\\
2 R_{XYWZ}  &=& -\frac{1}{z^0}\left(   g_{Z[X}g_{Y]W}+ \vec J_{XY}\cdot
\vec J_{ZW}- \vec J_{Z[X}\cdot \vec J_{Y]W}\right) + L_{ZW}{}^{AB}
\mathcal{W} _{ABCD}L_{XY}{}^{CD}.\nonumber
\end{eqnarray}
We raise and lower $A,B$ indices using $C_{AB}$ and $X,Y$ indices using
$g_{XY}$, see~(\ref{defgXY}).

For every point in $\{z^0,\,z^\a\}$, the $\{q^X\}$ subspace describes a
quaternionic-K{\"a}hler manifold. These quaternionic-K{\"a}hler manifolds are
related by coordinate redefinitions, $\SU(2)$ gauge transformations
and/or dilatations. Note that before gauge-fixing, all unhatted objects
\textit{a priori} are still dependent on both $z^0$, $z^\alpha$ and
$q^X$. For every gauge-fixing, eliminating the compensator fields $z^0,
z^\alpha$ in terms of constants (or functions of $q^X$), they become
geometrical objects on the quaternionic-K{\"a}hler manifold.

Note that in this section we only discussed the geometry related to the
hypermultiplets. The (very special) geometry related to the
vector-multiplets will be discussed in the next section when we perform
the gauge-fixing. The reason for this is that the discussion of the
vector-multiplet geometries requires the use of the equations of motion
and therefore cannot be discussed independently of the physical theory.

%%%%%%%%%%%%%%%%%%%%%%%%%%%%%%%%%%%%%%%%%%%%%%%%%%%%%%%%%%%%%%%%%%%%%%%%%%%%%%%
\section{Gauge-fixing\label{s:gf}}
%%%%%%%%%%%%%%%%%%%%%%%%%%%%%%%%%%%%%%%%%%%%%%%%%%%%%%%%%%%%%%%%%%%%%%%%%%%%%%%

The actions given in appendix~\ref{app:conf-matter} are invariant under
the full superconformal group. In order to break the symmetries that are
not present in the Poincar{\'e} algebra, we will impose the necessary gauge
conditions in the following subsections. This is analogous to the
corresponding steps in $N=2$ supergravity in $D=4$~\cite{deWit:1985px}
and in $D=6$~\cite{Bergshoeff:1986mz}.

\begin{figure}[tb]
\ifpdf
  \includegraphics{matter.pdf}
 \else
  \centerline{\epsfig{file=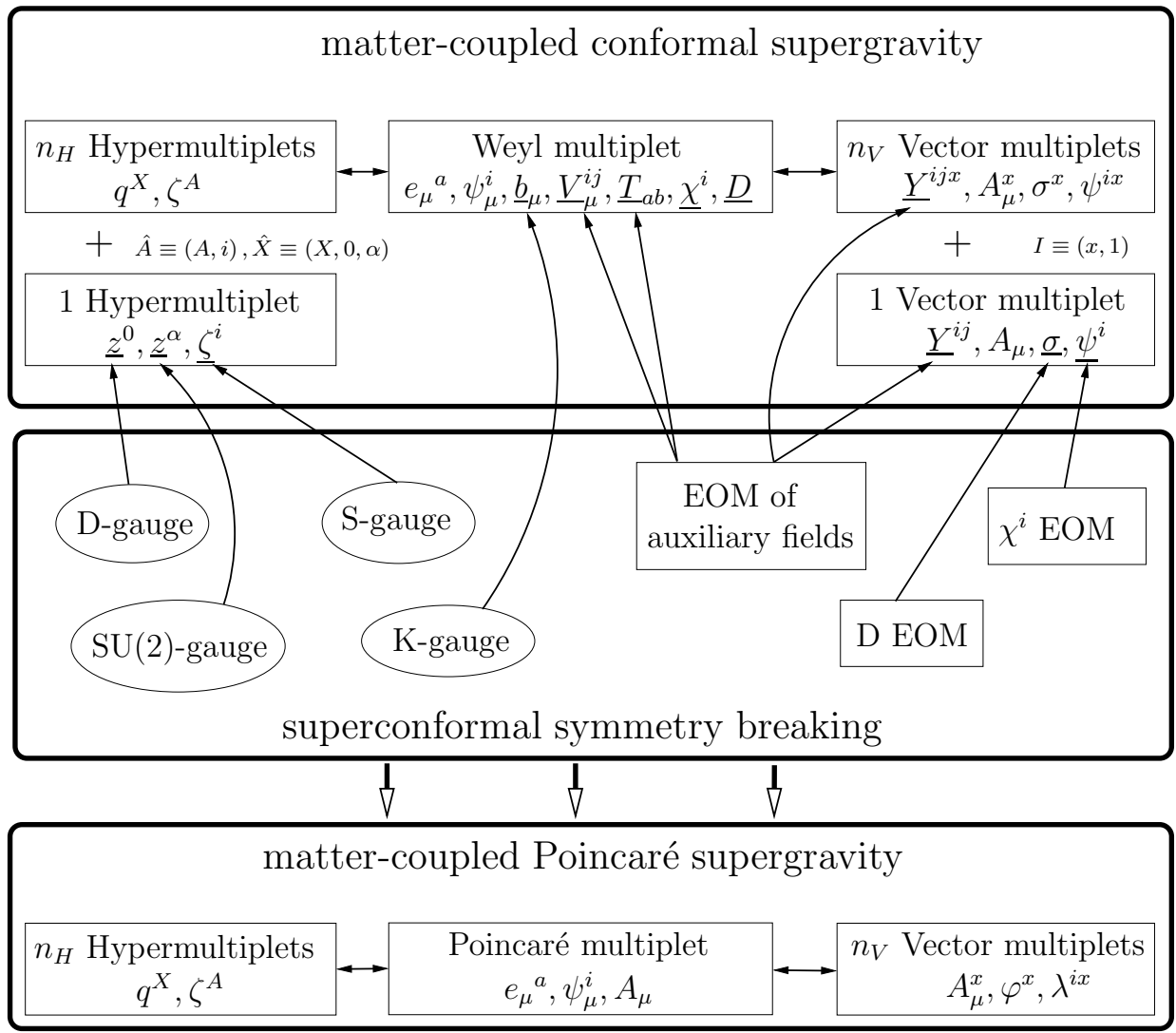,width=\textwidth}}
 \fi
 \caption{\it The gauge-fixing procedure: the underlined fields are
eliminated when passing to Poincar{\'e} SUGRA. The  arrows indicate how these
fields are eliminated: by gauge-fixing a symmetry or by applying an
equation of motion.}
 \label{fig:matter}
\end{figure}

Before carrying out all technicalities implied by the gauge-fixing
procedure, it is instructive to outline the steps we are going to follow.
Just like in the example of section~\ref{s:intr} the extra
(superconformal) symmetries can be removed with the help of compensating
multiplets. In our particular case, one hypermultiplet together with  one
vector multiplet will play the role of compensating multiplets. Our
strategy is illustrated in figure~\ref{fig:matter}. In this figure we
have summarized which fields are eliminated by gauge-fixing and/or
solving the equations of motion\footnote{The arrows show that one can
imagine another construction. Namely we may obtain conformally invariant
actions with consistent field equations by only using a compensating
vector multiplet and no compensating hypermultiplet. The compensating
vector multiplet is necessary to obtain consistent field equations for
the auxiliary $D$ and $\chi_i$. However, the compensating hypermultiplet
is only needed to break conformal invariance.}.

The field content of the matter-coupled conformal supergravity is given
by the Weyl multiplet, $n_H +1$ hypermultiplets and $n_V+1$ Yang-Mills
vector multiplets combined with $n_T$ tensor multiplets in a
vector-tensor multiplet. Note that in figure~\ref{fig:matter} we
represented only the independent fields of the Weyl multiplet. The
dependent fields $f_\mu{}^a$, $\hat{\omega}_\mu{}^{ab}$ and $\phi_\mu^i$
are expressed in terms of the former in the first step (see below). Note
that the $b_\mu$ field does not enter the action; it can therefore be set
to zero as a gauge condition for the special conformal (or
$K$-)transformations. There exist several auxiliary fields both in the
standard Weyl multiplet ($V_\mu^{ij}$ and $T_{ab}$) as in the vector
multiplets ($Y^{ij  I}$). They are eliminated by solving the
corresponding field equations.

The equations of motion of the $\chi^i$ and $D$ fields collaborate with
the $S$-gauge and $D$-gauge, respectively, to remove the fermionic
degrees of freedom of the compensating multiplets and two of their
scalars, i.e.~$\sigma$ and $z^0$. We see from the figure that all field
components of the compensating hypermultiplet are eliminated by the
gauge-fixing procedure. The only field component of the compensating
vector multiplet surviving the gauge-fixing procedure is the gauge
potential $A_\mu$, which contributes to the graviphoton field in the
Poincar{\'e} multiplet. As we eliminated the auxiliary fields $Y^{ij}$, the
vector multiplets, as well as their tensor multiplet companions, are
realized on-shell in the Poincar{\'e} theory. We thus end up with a
matter-coupled Poincar{\'e} supergravity theory containing, besides the
Poincar{\'e} multiplet, $n_V$ vector, $n_T$ tensor and $n_H$ hypermultiplets.
The geometry described by the moduli of the latter modifies during the
gauge-fixing according to our discussion in section~\ref{s:confgeo}. We
will see below that the vector scalars also parametrize  a particular
type of manifold at the Poincar{\'e} level, namely a very special real
manifold (see section~\ref{ss:hypersurfaces}).

\landscape
\endlandscape

One should keep in mind that during the gauge-fixing procedure the
definition of the covariant derivatives changes. Indeed, when passing
from a superconformal invariant theory to a super-Poincar{\'e} theory, the
remaining fields are chosen such that they do not transform under the
broken symmetries, e.g.~the scale symmetries. These scale symmetries
generated terms in the superconformal covariant derivative that are
absent in the Poincar{\'e} covariant derivatives. Another thing to keep in
mind has to do with the transformation rules. In the conformally
invariant theory, these transformation rules involve the parameters of
the conformal transformations. Due to the gauge-fixing conditions, these
parameters become dependent and are expressed in terms of the parameters
of the Poincar{\'e} theory through the so-called decomposition rules. The
super-Poincar{\'e} transformation rules are therefore inferred from the
superconformal ones after eliminating auxiliary fields and employing the
decomposition rules. This finishes our overview of the gauge-fixing
procedure. We now proceed with a more technical discussion of the same
procedure.

%%%%%%%%%%%%%%%%%%%%%%%%%%
\subsection{Preliminaries}
%%%%%%%%%%%%%%%%%%%%%%%%%%

The first step in the gauge-fixing process is the elimination of the
dependent gauge fields $\phi_\mu^i$ and $f_\mu{}^a$, associated with $S$-
and $K$-symmetries, respectively. Using the relations given
in~\cite[(3.11)]{Bergshoeff:2001hc} together with the definitions of the
supercovariant curvatures, we find the following expressions for these
gauge fields:
\begin{eqnarray}
f_a^{~a}
&=& \ft{1}{16} \left(- R(\hat\omega) - \ft13 \bar{\psi}_\rho\gamma^{\rho\mu\nu} {\cal D}_\mu \psi_\nu \right.\nonumber\\
&&+ \left.\ft13 \bar{\psi}_a^i\gamma ^{abc}\psi_b^jV_{cij} + 16 \bar{\psi}_a\gamma ^a\chi - 4 {\rm i}\bar{\psi}^a\psi^bT_{ab} + \ft43 {\rm i}\bar{\psi}^b\gamma _{abcd}\psi^aT^{cd} \right) , \nonumber\\
\hat\omega_\mu^{~ab}
&=& \omega_\mu^{~ab}(e) -\ft12\bar{\psi}^{[b}\gamma ^{a]}\psi_{\mu}-\ft14\bar{\psi}^b\gamma _\mu \psi^a + 2 e_\mu{}^{[a}b^{b]} , \nonumber\\
\phi^i_\mu &=& \ft12\,{\rm i} \gamma ^\nu {\cal D}_{[\mu}\psi^i_{\nu]} -
\ft{1}{12}\,{\rm i} \gamma _\mu{}^{\nu\rho} {\cal D}_{\nu}\psi^i_{\rho}
 - \ft12\,{\rm i} V_{[\mu}{}^{ij} \gamma ^\nu  \psi _{\nu ]j}+\ft1{12}\,{\rm i} V_{a}{}^{ij} \gamma _\mu{}^{ab} \psi _{bj} \nonumber\\
&& -\, T^a{}_\mu \psi_a^i - \ft13T^{ab}\gamma _{b\mu}\psi_a^i - \ft23
T_{b\mu}\gamma ^{ab}\psi_a^i - \ft13 T_{bc}\gamma ^{abc}{}_\mu \psi_a^i \nonumber\\
&& -\, \ft{1}{12}\,{\rm i} (\gamma ^{ab} \gamma _\mu  - \ft12 \gamma _\mu
\gamma ^{ab} ) b_a \psi _b^i , \label{transfDepF}
\end{eqnarray}
with
\begin{eqnarray} \label{Lor-covder}
{\cal D}_\mu &=&
\partial_\mu + \ft14 \hat\omega_\mu{}^{ab} \gamma _{ab} .
\end{eqnarray}
We only need the contracted version of $f_\mu{}^a$ since the other
components do not appear in the action or transformation rules. In order
to simplify the notation we choose to work with $\hat\omega_\m{}^{ab}$
instead of $\omega_\m{}^{ab}$ for the time being.

After writing out all covariant derivatives and dependent gauge fields,
the gauge field $b_\mu$ does not appear in the action. This can be
understood from the special conformal symmetry (or $K$-symmetry) of the
action. We will choose the conventional gauge choice for the
$K$-invariance, namely
\begin{equation}\label{eq:K}
K\mbox{-gauge:}\quad b_\mu=0.
\end{equation}

At this point we are left with one more gauge field corresponding to a
non-Poincar{\'e} symmetry: the $\SU(2)$ gauge field $V_\mu^{ij}$. Solving for
its equation of motion, corresponding to the
action~(\ref{eq:superconf-action}), yields the following expression:
\begin{eqnarray}
V_\mu^{ij} &=& \frac{9}{2 k^2}\Big( \hat g_{{\hat X}{\hat Y}}
\big(\partial_\mu q^{\hat X} + g A_\mu^I \hat{k}_I^{\hat X}\big)
\hat{k}^{ij{\hat Y}} + \frac12\,{\rm i} \hat{k}^{\hat X} {\hat f}_{{\hat
X}}^{(i{\hat A}} \bar{\zeta}_{\hat A} \gamma _{\mu\nu} \psi^{\nu j)} -
{\rm i} \hat{k}^{ij{\hat X}} {\hat f}_{k{\hat X}}^{\hat A} \bar
\zeta_{\hat A} \gamma _\nu\gamma _\mu\psi^{k\nu} \nonumber\\&& - \ft 12
C_{\widetilde I\widetilde J\widetilde K} \sigma^{\widetilde K}
\bar{\psi}^{i\widetilde I}\gamma _\mu \psi^{j\widetilde J} + \frac14\,
{\rm i} C_{\widetilde I\widetilde\, J\,\widetilde K}\sigma^{\widetilde
K}\sigma^{\widetilde I} \bar{\psi}^{i\widetilde J}\gamma _{\mu\nu}
\psi^{j\nu} \Big).
\end{eqnarray}

The action further contains four auxiliary matter fields: $D$, $T_{ab}$
and $\chi^i$ from the Weyl multiplet, and $Y^{\widetilde I}_{ij}$ from
the vector-tensor multiplet. Both $D$ and $\chi^i$ appear as Lagrange
multipliers in the action, leading to the following constraints,
\begin{eqnarray}
D &:& \qquad C - \ft13 k^2 = 0 , \qquad \mbox{with} \quad C \equiv
C_{\widetilde I\,\widetilde J\,\widetilde K} \sigma ^{\widetilde I}
\sigma ^{\widetilde J}\sigma ^{\widetilde K} , \quad
k^2= \hat{k}^{\hat X}g_{\hat{X}\hat{Y}}\hat{k}^{\hat Y}=-9z^0,\label{eq:D-EOM} \\
&& \nonumber\\
\chi^i &:& \qquad -8 {\rm i} C_{\widetilde I\widetilde J\widetilde
K}\sigma ^{\widetilde I}\sigma ^{\widetilde J}\psi^{\widetilde K}_i -
\ft43 \left( C -\ft13 k^2 \right) \gamma ^\mu \psi_{\mu i} +
\ft{16}{3}{\rm i} A^{\hat A}_i \zeta_{\hat A} = 0. \label{eq:chi-EOM}
\end{eqnarray}
Here, and for later purposes, we have introduced sections $A_i^{\hat A}$
that are defined by~\cite{deWit:1999fp}
\begin{equation}
A_i^{\hat A} =  {\hat k}^{\hat X} f_{i{\hat X}}^{\hat A}.
\end{equation}
The equations of motion for $Y^{\widetilde I}_{ij}$ and $T_{ab}$ are
given by
\begin{eqnarray}
&& Y^{ij\widetilde J} C_{\widetilde I\widetilde J\widetilde K} \sigma^{\widetilde K} = - g  \delta_{\widetilde I}^L \hat P_L^{ij} + \ft14\,{\rm i} C_{\widetilde I\widetilde J\widetilde K} \bar\psi^{i\widetilde J} \psi^{j\widetilde K} , \label{eq:Y-EOM} \\
&& T_{ab} = \frac{9}{64 k^2} \Big( 4 \sigma^{\widetilde
I}\sigma^{\widetilde J}\widehat{\mathcal H}^{\widetilde
K}_{ab}C_{\widetilde I\,\widetilde J\,\widetilde K} + \sigma^{\widetilde
I}\sigma^{\widetilde J}\bar{\psi}^{\widetilde
K}\gamma_{[a}\psi_{b]}C_{\widetilde I\,\widetilde J\,\widetilde K} +
\sigma^{\widetilde I}\sigma^{\widetilde J}\bar{\psi}^{\widetilde
K}{\gamma}_{abc}\psi^c C_{\widetilde I\,\widetilde J\,\widetilde K}
\label{eq:T-EOM}\\
&& \qquad\qquad +\, {\rm i}\sigma^{\widetilde I}\,\bar{\psi}^{\widetilde
J}{\gamma}_{ab} \psi^{\widetilde K}C_{\widetilde I\widetilde J\widetilde
K} + \ft23 \hat{k}^{\hat X} {\hat f}_{i{\hat X}}^{\hat A}
\bar{\zeta}_{\hat A} {\gamma}_{[a}\psi_{b]}^i + \ft23\hat{k}^{\hat X}
{\hat f}_{i{\hat X}}^{\hat A} \bar{\zeta}_{\hat A}
{\gamma}_{abc}\psi^{ic} + 2{\rm i} \bar{\zeta}_{\hat A}{\gamma}_{ab}
\zeta^{\hat A}\Big) .\nonumber
\end{eqnarray}
These equations have been simplified by using~(\ref{eq:D-EOM}).

%%%%%%%%%%%%%%%%%%%%%%%%%%%%%%%%%%%%%%%%%%%%%%%%%%
\subsection{Gauge choices and decomposition rules}
\label{ss:gaugechoice}
%%%%%%%%%%%%%%%%%%%%%%%%%%%%%%%%%%%%%%%%%%%%%%%%%%

Apart from the $K$-gauge (\ref{eq:K}) that we have already introduced to
fix the special conformal symmetry, we now choose gauges for the other
non-Poincar{\'e} (super)symmetries as well.

\paragraph{D-gauge.}

Demanding canonical factors for the Einstein-Hilbert and Rarita-Schwinger
kinetic terms in~(\ref{eq:superconf-action}), we have to impose the
following $D$-gauge:
\begin{equation}
 \label{eq:D-gauge}
D\mbox{-gauge:}\quad  \frac{1}{24} \left( C + k^2 \right) = -\frac{1}{2
\kappa^2} .
\end{equation}
where $\kappa$ has dimensions of [length]$^{3/2}$. If we combine the
D-gauge~(\ref{eq:D-gauge}) and D-EOM~(\ref{eq:D-EOM}) we obtain
\begin{equation}
\label{eq:hypersurfaces}
 k^2 = -\frac{9}{\kappa^2}, \qquad  C = -\frac{3}{\kappa^2}.
\end{equation}
In the light of (\ref{Killk})-(\ref{QK}) the first constraint implies
that
\begin{equation}\label{eq:z0}
z^0=\kappa^{-2},
\end{equation}
whereas the second constraint effectively eliminates one of the
vector-tensor scalars\footnote{The constraint (\ref{eq:z0}) implies that
the parameter $\nu$ defined in \cite{Bergshoeff:2002qk} is given by $\nu
= -\kappa^2$. This parameter also appeared in (\ref{equivariance}) but
from now on will not appear anymore in this paper.}.

\paragraph{S-gauge.}
\label{ss:S-gauge}

In the action~(\ref{eq:superconf-action}) there appear terms where
$\gamma ^\mu \phi_\mu= \ft14\,{\rm i}\gamma^{\mu \nu}\partial_\mu
\psi_\nu  +$ (non-derivative terms) is multiplied by hyperino and gaugino
fields. These terms imply a mixing of the kinetic terms of the gravitino
with the hyperino and gaugino fields. A suitable $S$-gauge can eliminate
this mixing: we put the coefficient of the above expression equal to
zero:
\begin{equation}
  S\mbox{-gauge:}\quad C_{\widetilde I\,\widetilde J\,\widetilde K}
  \sigma^{\widetilde I} \sigma^{\widetilde J} \psi_i^{\widetilde K} +
  2 A_i^{\hat{A}} \zeta_{\hat{A}} =0.
 \label{eq:S-gauge}
\end{equation}
Combining this with the $\chi $ field equation~(\ref{eq:chi-EOM}) leads
to
\begin{equation}
  C_{\widetilde I\,\widetilde J\,\widetilde K}
  \sigma^{\widetilde I} \sigma^{\widetilde J} \psi_i^{\widetilde K} =0,\qquad
   A_i^{\hat{A}} \zeta_{\hat{A}}=0.
 \label{eq:S-gaugesplit}
\end{equation}
In our coordinate basis, we obtain the following expression for the
sections $A^i_{\hat A}$:
\begin{equation}
\label{eq:genSU2gauge}
 A^i_{\hat A} \equiv \varepsilon^{ij}\hat{k}_{\hat X} \hat f^{\hat X}_{j\hat{A}} = -3\varepsilon^{ij} \hat{f}^0_{j\hat{A}}
 = -3\,{\rm i} \sqrt\frac{z^0}{2} \delta_{\hat A}^i .
\end{equation}
Therefore, our choice of coordinates on the hyper-K{\"a}hler manifold is
consistent with the fact that the hyperinos of the compensating
hypermultiplet carry no physical degree of freedom:
\begin{equation}
\label{eq:zetai} \zeta^i = 0.
\end{equation}

\paragraph{$\SU(2)$-gauge.}

The gauge for dilatations was chosen such that $z^0=\kappa^{-2}$.
Similarly we may also choose a gauge for $\SU(2)$. Such a gauge would be
a specific point in the 3-dimensional space of the $z^\alpha $. In
principle, we could thus choose $z^\alpha =z_0^\alpha (q)$ for any
function $z_0^\alpha (q)$, but we will restrict ourselves here to
constants $z_0^\alpha$:
\begin{equation}
  \mbox{SU(2)-gauge:} \qquad z^\alpha = z_0^\alpha .
 \label{SU2gauge}
\end{equation}

\paragraph{Decomposition rules.}

As a consequence of the gauge choices, the corresponding transformation
parameters can be expressed in terms of the independent ones by the
so-called decomposition rules. For example, the requirement that the
$K$-gauge (\ref{eq:K}) should be invariant under the most general
superconformal transformation, i.e. $\delta b_\mu=0$, leads to the
following decomposition rule for $\Lambda_K^a$:
\begin{equation}
\Lambda_K^a = -\ft12 e^{\mu a} \left( \partial_\mu \Lambda_D + \ft12\,
{\rm i} \bar\e\phi_\mu - 2 \bar\e{\gamma}_\mu\chi + \ft{1}{2}\,{\rm i}
\bar\eta\psi_\mu \right).
\end{equation}
Similarly, demanding $\delta z^0=0$ yields
\begin{equation}
  \Lambda_D = 0.
\end{equation}
The decomposition rule for $\eta^i$ can be found by varying the S-gauge and demanding that
\begin{equation}
\delta \left( C_{\widetilde I\,\widetilde J\,\widetilde
K}\sigma^{\widetilde I}\sigma^{\widetilde J}\psi^{i\widetilde K} \right)
= 0. \label{DemanddelC0}
\end{equation}
We find
\begin{eqnarray}
\label{eq:eta} \kappa ^{-2}\eta^i
 &=& - \ft{1}{12} C_{\widetilde I\,\widetilde J\,\widetilde K} \sigma^{\widetilde I}
  \sigma^{\widetilde J} \gamma  \cdot {\widehat{\mathcal H}}^{\widetilde K} \e^i
  + \ft{1}{3} g \sigma^I P_I^{ij} \e_j
 + \ft{1}{32 } \,{\rm i} \gamma ^{ab} \e^i \bar\zeta_A {\gamma}_{ab} \zeta^A \nonumber\\
&& +\, \ft{1}{16} \,{\rm i} C_{\widetilde I\,\widetilde J\,\widetilde K}
\sigma^{\widetilde I} \Big(\gamma^a \e_j \bar\psi^{i\widetilde J}
{\gamma}_a\psi^{j\widetilde K} - \ft{1}{16} \gamma ^{ab} \e^i
\bar\psi^{\widetilde J} {\gamma}_{ab}\psi^{\widetilde K}\Big) .
\end{eqnarray}
The $\SU(2)$ decomposition rule can be found by requiring that $\delta
z^{\alpha} = 0$:
\begin{eqnarray}
\vec \Lambda_{\SU(2)} &=& \vec \omega_X (\delta _Q + \delta _G) q^X + g
\Lambda_G^I \vec P_I.
\end{eqnarray}

%%%%%%%%%%%%%%%%%%%%%%%%%%
\subsection{Hypersurfaces\label{ss:hypersurfaces}}
%%%%%%%%%%%%%%%%%%%%%%%%%

We now discuss the geometry for the vector-multiplet scalars. These arise
as a consequence of the gauge-fixing procedure. In order to get a
standard normalization, we rescale the $C_{\widetilde I\,\widetilde
J\,\widetilde K}$ symbol and the vector multiplet scalars as follows:
\begin{equation}
\sigma^{\widetilde I} \equiv \sqrt{\frac{3}{2\kappa^2}} h^{\widetilde I},
\qquad C_{\widetilde I\,\widetilde J\,\widetilde K} \equiv -
2\sqrt{\frac{2\kappa^2}{3}} {\cal C}_{\widetilde I\,\widetilde
J\,\widetilde K},
\end{equation}
such that
\begin{equation}
{\cal C}_{\widetilde I\,\widetilde J\,\widetilde K} h^{\widetilde I}
h^{\widetilde J} h^{\widetilde K} = 1. \label{eq:hypersurface_vector}
\end{equation}
The constraint (\ref{eq:hypersurface_vector}) defines an $(n_V +
n_T)$-dimensional hypersurface of scalars $\phi^x$ called a `very special
real' manifold, embedded into a $(n_V + n_T +1)$-dimensional space
spanned by the scalars $h^{\widetilde I}(\phi)$.

The metric on the embedding $h^{\widetilde I}$-manifold can be determined
by substituting the equation of motion for $T_{ab}$ (\ref{eq:T-EOM}) back
into the action, and defining the kinetic term for the vectors/tensors as
\begin{equation}
 {\mathcal L}_{\rm kin, Vector-\!Tensor} = - \ft14 a_{\widetilde I\,\widetilde J} \widehat{\mathcal H}_{\mu\nu}^{\widetilde I} \widehat{\mathcal H}^{\mu\nu\widetilde J}.
\end{equation}
We find
\begin{eqnarray}
a_{\widetilde I\,\widetilde J} &=& -2 {\cal C}_{\widetilde I\,\widetilde
J\widetilde K} h^{\widetilde K} + 3 h_{\widetilde I} h_{\widetilde J},
\label{eq:a_IJ-identity}
\end{eqnarray}
where
\begin{eqnarray} \label{eq:normalized_hypersurface_vector}
  h_{\widetilde I} &\equiv& a_{\widetilde I\,\widetilde J} h^{\widetilde J} = {\cal C}_{\widetilde I\,\widetilde J\,\widetilde K} h^{\widetilde J} h^{\widetilde K} \qquad \Rightarrow \qquad h_{\widetilde I} h^{\widetilde I} = 1 .
\end{eqnarray}
In the following we will assume that $a_{\widetilde I\widetilde J}$ is
invertible; this enables us to solve~(\ref{eq:Y-EOM}) for $Y^{\widetilde
I ij}$. Following~\cite{Gunaydin:1984bi}, we introduce
\begin{equation}
 h_x^{\widetilde I} \equiv -\sqrt{\frac{3}{2\kappa^2}} h^{\widetilde I}_{,x}(\phi) \quad \rightarrow \quad h_{\widetilde Ix} \equiv a_{\widetilde I\,\widetilde J} h^{\widetilde J}_x(\phi) = \sqrt{\frac{3}{2\kappa^2}} h_{\widetilde I,x}(\phi) .
\end{equation}
The metric on the manifold spanned by the scalars $\phi^x$ is the
pull-back on the hypersurface of the metric $a_{\widetilde I \widetilde
J}$ on the embedding space, i.e.
\begin{equation}
  g_{xy} = h^{\widetilde I}_x h^{\widetilde J}_y
a_{\widetilde I\widetilde J}.
\end{equation}
We furthermore define
\begin{equation}
h^x_{\widetilde I} \equiv g^{xy} h_{\widetilde I y}.
\end{equation}
Several useful identities that were already discovered
in~\cite{Gunaydin:1984bi} are summarized in
appendix~\ref{app:veryspecial}.

The gauginos $\psi^{\widetilde I}$ are constrained fields, due to the
$S$-gauge. In order to translate these to $(n_V + n_T)$ unconstrained
gauginos, we introduce $\lambda^{i\,x}$, which transform as vectors in
the tangent space on the hypersurface. As we will see later, a convenient
choice is given by (for agreement with the
literature~\cite{Ceresole:2000jd})\footnote{We avoid here the
introduction of a local basis for the fermions indicated by indices
$\tilde a$ in~\cite{Ceresole:2000jd}.}
\begin{equation}
\label{eq:lambda} \lambda^{i x} \equiv - h^x_{\widetilde I}
\psi^{i\widetilde I}, \qquad \psi^{i\widetilde I} = - h^{\widetilde I}_x
\lambda^{i x} .
\end{equation}
Note that this choice for $\psi^{i\widetilde I}$ indeed solves the
S-gauge~(\ref{eq:S-gaugesplit}).

%%%%%%%%%%%%%%%%%%%%%%%%%%%%%%%%%%%%%%%%%%%%%%%%%%%%%%%%%%%%%%%%%%%%%%%%%%%%%%%
\section{Results}\label{s:results}
%%%%%%%%%%%%%%%%%%%%%%%%%%%%%%%%%%%%%%%%%%%%%%%%%%%%%%%%%%%%%%%%%%%%%%%%%%%%%%%

After applying the steps outlined in the previous section, i.e.~using a
special coordinate basis, substituting the expressions for the dependent
gauge fields and matter fields and `reducing' the objects on the
hyper-K{\"a}hler manifold to the quaternionic-K{\"a}hler manifold, we obtain the
$N=2$ super-Poincar{\'e} action.

We give in this section the full action for a number of vector multiplets
(indices $I$), tensor multiplets (indices $M$, and together with the
vector multiplets indicated as $\tilde I$, and their unconstrained fields
with $x$) and hypermultiplets (indices $X$ for the scalars and $A$ for
the spinors). The couplings of the vector and tensor multiplets are
determined by the constants ${\cal C}_{\tilde I\tilde J\tilde K}$, a
symplectic metric $\Omega_{MN}$ and the transformation matrices
$t_{I\tilde J}{}^{\tilde K}$ related by~(\ref{COmegarelations}), see
also~(\ref{matricest}). The related quantities are defined in
section~\ref{ss:hypersurfaces}.

We define the supercovariant field strengths $\widehat F^I_{ab}$ and a
tensor field $\widetilde B^M_{ab}$ such that
\begin{eqnarray}
&&\widehat{\mathcal{H}}^{\widetilde{I}}_{ab}=\left(\widehat F^I_{ab},
B^M_{ab} \right) =\left(F_{ab}^I-\bar{\psi}_{[a}\gamma_{b]}\psi^I
+\frac{\sqrt{6}}{4\kappa}\,{\rm i} \bar{\psi}_a\psi_b
h^I,\widetilde{B}_{ab}^M-\bar{\psi}_{[a}\gamma_{b]}\psi^M+
\frac{\sqrt{6}}{4\kappa} \,{\rm i} \bar{\psi}_a\psi_b h^M\right),\nonumber\\
&&H_{ab}^{\widetilde{I}} \equiv
\left(F_{ab}^I,\widetilde{B}_{ab}^M\right), \qquad F_{\mu\nu}^I \equiv  2
\partial_{[\mu} A_{\nu]}^I + g f_{JK}{}^I A_\mu^J A_\nu^K.
\end{eqnarray}
The $B^M_{ab}$ transforms covariantly, as does $\hat{F}_{ab}$, while the
action gets a simpler form using $F_{ab}$ and $\widetilde B_{ab}$.

The hypermultiplets are completely characterized in terms of the
vielbeins $f_X^{iA}$, that determine complex structures, $\USp(2n_H)$ and
$\SU(2)$ connections. They transform in general under the gauge group of
the vector multiplets. The Killing vectors $k_I^X$ determine $t_{IA}{}^B$
and are restricted by (\ref{isometries}). They determine the moment maps
by~(\ref{momentmap}). As mentioned in section~\ref{s:confgeo}, the moment
maps can also exist without a quaternionic-K{\"a}hler manifold ($n_H=0$), in
which case they are the constant `FI terms'. These are possible for two
cases. First, in the case where the gauge group contains an $\SU(2)$
factor, we can have
\begin{equation}
  \vec P_I= \vec e_I \xi ,
 \label{SU2FI}
\end{equation}
where $\xi$ is an arbitrary constant, and $\vec e_I$ are constants that
are nonzero only for $I$ in the range of the $\SU(2)$ factor and satisfy
\begin{equation}
  \vec e_I\times \vec e_J= f_{IJ}{}^K \vec e_K,
 \label{vece}
\end{equation}
in order that~(\ref{equivariance}) is verified.

The second case is $\U(1)$ FI terms. In this case
\begin{equation}
   \vec P_I= \vec e\, \xi_I ,
 \label{U1FI}
\end{equation}
where $\vec e$ is an arbitrary vector in $\SU(2)$ space and $\xi_I$ are
constants for the $I$ corresponding to $\U(1)$ factors in the gauge
group.

To be able to write down the potential and the supersymmetry
transformation rules in an elegant fashion, we define
\begin{equation}
\begin{array}{ll} W^x \equiv \frac{\sqrt6}{4}\kappa h^I K^x_I = -\frac{3}{4}
t_{J{\widetilde I}}{}^{\widetilde P} h^J h^{\widetilde I} h^x_{\widetilde
P} , \qquad & K^x_I \equiv -\frac1\kappa \sqrt\frac32 t_{I{\widetilde
J}}{}^{\widetilde K} h^{\widetilde J} h^x_{\widetilde K} =-\frac1\kappa
\sqrt\frac32 t_{I{\widetilde J}}{}^{\widetilde K} h^{{\widetilde J}x}
h_{\widetilde K}
,  \\
\vec P \equiv \kappa ^2 h^I \vec P_I , \qquad & \vec P_x \equiv \kappa ^2
h_x^I \vec P_I , \\ {\mathcal N}^{iA} \equiv \frac{\sqrt6}{4}\kappa
h^I k_I^X f_X^{iA},&  \\
T_{xyz} \equiv
\mathcal{C}_{\widetilde{I}\widetilde{J}\widetilde{K}}h^{\widetilde{I}}_xh^{\widetilde{J}}_yh^{\widetilde{K}}_z,
& \Gamma_{xy}^w = h^w_{\widetilde I} h^{\widetilde I}_{x,y} + \kappa
\sqrt{\frac23} T_{xyz} g^{zw},
\end{array}
\end{equation}
the latter being the Levi-Civita connection of $g_{xy}$.

The covariant derivatives now read
\begin{eqnarray}
\mathcal{D}_\mu h^{\widetilde I} &=& \partial_\mu h^{\widetilde I} + g
t_{J\widetilde K}{}^{\widetilde I} A_\mu^J h^{\widetilde K}
 =-\sqrt{\ft23}\kappa h_x^{\widetilde{I}}\left( \partial_\mu \phi^x+gK_J^xA_\mu^J\right) =-\sqrt{\ft23}\kappa h_x^{\widetilde{I}}\mathcal{D}_\mu \phi^x  , \nonumber\\
{\cal D}_\mu q^{X}
&=& \partial_\mu q^{X}+ g A_\mu^I k_I^{X}, \nonumber\\
\mathcal{D}_\mu \lambda^{xi}&=&\partial_\mu\lambda^{xi}+\partial_\mu
\phi^y \Gamma_{yz}^x\lambda^{zi}+\ft14 \omega_\mu{}^{ab}\gamma
_{ab}\lambda^{xi}\nonumber\\&& +\partial_\mu q^X
\omega_{Xj}{}^i\lambda^{xj}
- g\kappa ^2 A^I_\mu P_{I}{}^{ij}\lambda^{x}_j +g A_\mu^I K_I^{x;y} \lambda_y^i,\nonumber\\
\mathcal{D}_\mu \zeta^{A}
&=& \partial_\mu \zeta^{A} + \partial_\mu q^{X} \omega_{{X}{B}}{}^{A} \zeta^{B} + \ft14 \omega_\mu{}^{bc} {\gamma}_{bc} \zeta^{ A} + g A^I_\mu t_{IB}{}^{A} \zeta^{B}, \nonumber\\
\mathcal{D}_\mu \psi_{\nu} ^i&=&\left(\partial_\mu +\ft14
\omega_\mu{}^{ab}\gamma_{ab}\right)\psi_{\nu }^i-\partial_\mu
q^X\omega_{X}{}^{ij}\psi_{\nu j}-g\kappa ^2 A_\mu^IP_{I}{}^{ij}\psi_{\nu
j} .
\end{eqnarray}
Here $K_I^{x;y}$ stands for the covariant derivative, where the index is
raised with the inverse metric $g^{xy}$. We choose to extract the
fermionic terms from the spin connection, using $\omega_{\mu a}{}^b$
instead of $\hat{\omega}_{\mu a}{}^b$ in the covariant derivatives and
the Ricci scalar, unless otherwise mentioned.

Performing all the steps of the conformal programme we find the following
action:
{\allowdisplaybreaks
\begin{eqnarray}\label{final}
e^{-1}\mathcal{L}&=&
%Bosonic Kinetic terms
\ft1{2\kappa^2} R(\omega) -\ft14a_{\widetilde{I}\widetilde{J}}
\widehat{\mathcal{H}}^{\widetilde{I}}_{\mu
\nu}\widehat{\mathcal{H}}^{\widetilde{J}\mu \nu} -\ft12
g_{xy}\mathcal{D}_\mu \phi^x\mathcal{D}^\mu \phi^y
-\ft1{2}g_{XY}\mathcal{D}_\mu q^{X}\mathcal{D}^\mu q^{Y}\nonumber\\&& +\,
\frac {1}{16 g} e^{-1} \varepsilon ^{\mu \nu \rho \sigma \tau}
\Omega_{MN} \widetilde{B}_{\mu \nu}^M \left(\partial_\rho
\widetilde{B}_{\sigma \tau}^N+2g t_{IJ}{}^NA_\rho ^I F_{\sigma \tau}^J
   + g t_{IP}{}^N A_\rho^I \widetilde{B}_{\sigma\tau}^P\right) \nonumber\\&&
%Fermionic Kinetic terms
-\,\ft1{2\kappa ^2}\bar{\psi}_\rho \gamma^{\rho \mu \nu}\mathcal{D}_\mu
\psi_\nu -\ft12 \bar \lambda_x\slashed{\mathcal{D}}\lambda ^x
-\bar{\zeta}^A\slashed{\mathcal{D}}\zeta_A\nonumber\\&&
%Potential
+\ft{g^2}{\kappa^4}\left( 4\vec{P}\cdot \vec{P}
-2\vec{P}^x\cdot\vec{P}_x- 2 W_x W^x
-2\mathcal{N}_{iA}\mathcal{N}^{iA}\right)\nonumber\\&&
%Chern-Simons terms
+\,\ft{\kappa}{6\sqrt{6}} e^{-1} \varepsilon ^{\mu \nu \lambda \rho \sigma } {\cal C}_{IJK} A_\mu^I \left[ F_{\nu\lambda }^J F_{\rho \sigma }^K + f_{FG}{}^J A_\nu^F A_\lambda ^G \left(- \ft12 g F_{\rho \sigma }^K + \ft1{10} g^2 f_{HL}{}^K A_\rho ^H A_\sigma^L \right)\right] \nonumber\\
& & -\, \ft{1}{8} e^{-1} \varepsilon^{\mu\nu\lambda\rho\sigma}
\Omega_{MN} t_{IK} {}^M t_{FG}{}^N A_{\mu}^I A_\nu^F A_\lambda^G
\left(-\ft 12 g  F_{\rho\sigma}^K + \ft{1}{10} g^2 f_{HL}{}^K A_\rho^H
A_\sigma^L \right)\nonumber\\&&
%Quadratic fermions
+\,\ft14 h_{\widetilde{I}x} H_{bc}^{\widetilde{I}}
\bar{\psi}_a\gamma^{abc}\lambda^x-\ft{\sqrt{6}}{16\kappa }\,{\rm i}
h_{\widetilde{I}}
H^{cd\widetilde{I}}\bar{\psi}^a\gamma_{abcd}\psi^b+\ft14\sqrt{\ft23}
\kappa {\rm i} \left(\ft14g_{xy}h_{\widetilde{I}}+ T_{xyz}
h_{\widetilde{I}}^z \right) \lambda^x \gamma \cdot
H^{\widetilde{I}}\lambda^y\nonumber\\&& +\ft18 \sqrt{6}\,{\rm i} \kappa
h_{\widetilde{I}}\bar{\zeta}_A\gamma \cdot H^{\widetilde{I}}\zeta^A
+\ft12 {\rm i} \bar{\psi}_a\slashed{\mathcal{D}} \phi^x \gamma^a\lambda_x
+{\rm i} \bar{\zeta}_A \gamma^a\slashed{\mathcal{D}}q^X\psi_a^if_{iX}^A
\nonumber\\&&
 +\,g\Big[-\sqrt{\ft32}\ft1{\kappa}\,{\rm i} h^It_{IB}{}^A\bar{\zeta}_A\zeta^B +2{\rm i}
k_I^Xf_{iX}^Ah^I_x \bar{\zeta}_A\lambda^{ix} \nonumber\\&&
\,\,-\sqrt{\ft23}\ft1{\kappa} \,{\rm i} \left( \ft14g_{xy} P_{ij}+T_{xyz}
P_{ij}^z\right) \bar{\lambda}^{ix}\lambda^{jy}
+\ft1{\kappa}\bar{\lambda}^x\lambda^y\left( \ft12\,{\rm i} h^I_x
K_{Iy}-\ft23
\partial_xW_y \right) \nonumber\\&&
 \,\, -\ft2{\kappa^2}{\cal
N}_i^A\bar{\zeta}_A\gamma^a\psi_a^i +\ft1{\kappa
^2}\bar{\psi}_a^i\gamma^a\lambda^{jx}\left( P_{xij}+\varepsilon
_{ij}W_x\right) +\sqrt{\ft38}\ft1{\kappa^3}\,{\rm i}
P_{ij}\bar{\psi}_a^i\gamma^{ab}\psi_b^j \Big] \nonumber\\&&
%Quartic fermions
-\,\ft{1}{32}\bar{\psi}_a^i \psi^{ja}\bar{\lambda}_i^x\lambda_{jx}
-\ft{1}{32}\bar{\psi}_a^i
\gamma_b\psi^{ja}\bar{\lambda}_i^x\gamma^b\lambda_{jx}
-\ft{1}{128}\bar{\psi}_a \gamma_{bc}\psi^{a}
\bar{\lambda}^x\gamma^{bc}\lambda_x \nonumber\\&&
-\,\ft{1}{16}\bar{\psi}_a^i\gamma^{ab}
\psi_b^j\bar{\lambda}^x_i\lambda_{jx}
-\ft{1}{32}\bar{\psi}^{ai}\gamma^{bc}
\psi^{dj}\bar{\lambda}^x_i\gamma_{abcd}\lambda_{jx} +\ft{1}{8\kappa
^2}\bar{\psi}_a\gamma_b\psi^b\bar{\psi}^a\gamma_c\psi^c\nonumber\\&&
-\,\ft{1}{16\kappa^2}\bar{\psi}_a\gamma_b\psi_c\bar{\psi}^a\gamma^c\psi^b
-\ft{1}{32\kappa^2}\bar{\psi}_a\gamma_b\psi_c\bar{\psi}^a\gamma^b\psi^c
+\ft{1}{32\kappa^2}\bar{\psi}_a\psi_b\bar{\psi}_c\gamma^{abcd}\psi_d
\nonumber\\
&&
-\,\ft{1}{16}\bar{\psi}^a\gamma^b\psi^c\bar{\zeta}_A\gamma_{abc}\zeta^A
 +\ft1{16}\bar{\psi}_a\gamma^{bc}\psi^a\bar{\zeta}_A\gamma_{bc}\zeta^A
-\ft1{16}\bar{\psi}^a\psi^b\bar{\zeta}_A\gamma_{ab}\zeta^A \nonumber\\&&
+\,\ft{\kappa}{6}\sqrt{\ft23}\,{\rm i} T_{xyz}\left(
\bar{\psi}_a\gamma_b\lambda^x\bar{\lambda}^y\gamma^{ab}\lambda^z
 +\bar{\psi}_a^i\gamma^a\lambda^{jx}\bar{\lambda}_i^y\lambda_j^z \right)
\nonumber\\&& +\,\ft{\kappa^2}{32}\bar{\lambda}^x\gamma_{ab}\lambda_x
\bar{\zeta}_A\gamma^{ab}\zeta^A
  -\ft{\kappa^2}{16}\bar{\lambda}^{ix}\gamma_a \lambda^j_x\bar{\lambda}^y_i
  \gamma^a\lambda_{jy}\nonumber\\&&
+\,\ft{\kappa^2}{128}\bar{\lambda}^x\gamma_{ab}\lambda_x\bar{\lambda}^y\gamma^{ab}\lambda_y
+\ft{\kappa^2}6g^{zt}T_{xyz}T_{tvw}
\bar{\lambda}^{ix}\lambda^{jy}\bar{\lambda}_i^v\lambda^w_j
-\ft{\kappa^2}{48}\bar{\lambda}^{ix}\lambda^j_x\bar{\lambda}_i^y\lambda_{jy}
\nonumber\\&&
+\,\ft{\kappa^2}{32}\bar{\zeta}_A\gamma_{ab}\zeta^A\bar{\zeta}_B\gamma^{ab}\zeta^B
-\ft1{4}\mathcal{W}_{ABCD}\bar{\zeta}^A\zeta^B\bar{\zeta}^C\zeta^D.
\end{eqnarray}
}

This action admits the following $N=2$ supersymmetry:
\begin{eqnarray}
\delta  e_\mu{}^a &=& \ft12 \bar\e \gamma^a \psi_\mu ,\nonumber\\
\delta \psi_\mu^i&=&D_\mu(\hat\omega)\e^i + \ft{{\rm i}\kappa} {4\sqrt6}
h_{\widetilde I} {\widehat{\mathcal H}}^{\widetilde I\nu \rho}
 ({\gamma}_{\mu\nu \rho} - 4 g_{\mu \nu} {\gamma}_\rho) \e^i + \delta q^X \omega_X^{ij} \psi_{\mu j} - \ft{1}{\kappa\sqrt6} {\rm i} g P^{ij}\gamma_\mu  \e_j\nonumber\\
&&-\,\ft{\kappa^2}{6} \e_j \bar\lambda^{ix} \gamma_\mu \lambda^j_x
+\ft{\kappa^2}{12}\gamma_{\mu \nu} \e_j \bar\lambda^{ix} \gamma^\nu
\lambda^j_x
 -\ft{\kappa^2}{48}\gamma_{\mu \nu \rho } \e_j \bar\lambda^{ix} \gamma^{\nu \rho} \lambda^j_x
 +\ft{\kappa^2}{12} \gamma^\nu  \e_j\bar\lambda^{ix} \gamma_{\mu \nu } \lambda^j_x\nonumber\\
&& + \ft{\kappa ^2}{16} {\gamma}_{\mu \nu \rho } \e^i\bar\zeta_A \gamma ^{\nu \rho } \zeta^A  ,\nonumber\\
\delta  h^{\widetilde I}&=& -\ft{\kappa}{\sqrt6} {\rm i}\bar\e\lambda^x
h_x^{\widetilde I}\quad ,\qquad
\delta\phi^x= \ft{1}{2} {\rm i}\bar\e\lambda^x,\nonumber\\
\delta  A_\mu^I&=&\vartheta_\mu^I,\nonumber\\
\delta \widetilde{B}^M_{\mu \nu}&=&2\mathcal{D}_{[\mu}\vartheta^M_{\nu]}
-\sqrt{6}\frac{g}{\kappa } \bar{\epsilon}\gamma_{[\mu}\psi_{\nu]}h_N
\Omega^{MN}
-{\rm i} g \bar{\epsilon}\gamma_{\mu \nu}\lambda^x h_{xN}\Omega^{MN},\nonumber\\
\delta  \lambda^{xi}
&=& - \ft{{\rm i}}{2} \widehat{\slashed{\mathcal D}} \phi^x \e^i - \delta\phi^y \Gamma_{yz}^x\lambda^{zi} + \delta q^X \omega_X{}^{ij} \lambda_j^x + \ft14 \gamma\cdot\widehat{\mathcal H}^{\widetilde I} h^x_{\widetilde I} \e^i \nonumber\\
&& \ft{\kappa }{4\sqrt6} T^{xyz} \left[ 3 \epsilon_j\bar\lambda_y^i
\lambda_z^j -  \gamma ^\mu\epsilon_j\bar\lambda_y^i {\gamma}_\mu
\lambda_z^j
- \ft12 \gamma ^{\mu\nu}\epsilon_j\bar\lambda_y^i {\gamma}_{\mu\nu} \lambda_z^j  \right]  - \ft{1}{\kappa^2} g P^{x\, ij} \e_j + \ft{1}{\kappa^2} g W^x \e^i ,\nonumber\\
\delta  q^X &=& -{\rm i}\bar\e^i\zeta^A f^X_{iA} ,\nonumber\\
\delta  \zeta^A&=& \ft12 \,{\rm i} \gamma^\mu \widehat{\mathcal D}_\mu
q^X f_X^{iA} \e_i - \delta  q^X \omega_{XB}{}^A \zeta^B +
\ft{1}{\kappa^2} g {\mathcal N}_i^A \e^i  ,
\end{eqnarray}
where $\Omega ^{MN}$ is minus the inverse of $\Omega_{MN}$ in the sense
that $\Omega_{MP}\Omega ^{NP}=\delta_M{}^N$. We also denoted
\begin{eqnarray}
 \vartheta_\mu^{\widetilde I} &=& -\ft12 \bar{\e} \gamma_\mu \lambda^x h_x^{\widetilde
I} - \ft{\sqrt6}{4\kappa } \,{\rm i} h^{\widetilde I} \bar\e \psi_\mu, \nonumber\\
 \mathcal{D}_{[\mu} \vartheta^{\widetilde{I}}_{\nu]}&=&\partial_{[\mu} \vartheta^{\widetilde{I}}_{\nu]}+gA_{[\mu}^Jt_{J\widetilde{K}}{}^{\widetilde{I}}\vartheta^{\widetilde{K}}_{\nu]}, \nonumber\\
\widehat{\mathcal D}_\mu q^X &=& \partial_\mu q^X + g A_{\mu}^I k_I^X
+ {\rm i} \bar\psi_\mu^j \zeta^B f_{jB}^X\nonumber\\
\widehat{\mathcal D}_\mu \phi^x &=& \partial_\mu \phi^x + g A_\mu^I K^x_I
- \ft{1}{2}\,{\rm i}
\bar\psi_\mu \lambda^x,\nonumber\\
D_\mu(\hat\omega) \e^i &=& {\mathcal D}_\mu(\hat\omega) \e^i -
\partial_\mu q^X \omega_X^{ij}\e_j - g\kappa^2 A_\mu^I P_I^{ij}\e_j,
\end{eqnarray}
where ${\mathcal D}_\mu(\hat\omega)$ is defined as in (\ref{Lor-covder}).

%%%%%%%%%%%%%%%%%%%%%%%%%%%%%%%%%%%%%%%%%%%%%%%%%%%%%%%%%%%%%%%%%%%%%%%%%%%%%%%
\section{Comparison with earlier papers}\label{s:earlier}
%%%%%%%%%%%%%%%%%%%%%%%%%%%%%%%%%%%%%%%%%%%%%%%%%%%%%%%%%%%%%%%%%%%%%%%%%%%%%%%
In this section, we compare our results with the literature, and
especially with \cite{Ceresole:2000jd} (CD) and
\cite{Gunaydin:1984bi,Gunaydin:1985ak,Gunaydin:1999zx,Gunaydin:2000xk,Gunaydin:2000ph}
(GZ)\footnote{There is an `alternative $N=2$
supergravity'~\cite{Nishino:2001ji}, which is very different from the
theory described here.}. To compare with these papers we put $\kappa =1$.
A notational difference with CD is that we have $ \gamma^{abcde}= {\rm i}
\varepsilon^{abcde} $, while CD uses $ \gamma^{abcde}= -{\rm i}
\varepsilon^{abcde}$. Another difference is in the symplectic metric
$\Omega_{MN}$. Our $\Omega_{MN}$ is 4 times the one in CD and in GZ.
Furthermore they use $\Omega_{MN}\Omega^{NP}=\delta_M^P$ and our
$t_{I\tilde J}{}^{\tilde K}$ is denoted as $\Lambda_{I\tilde J}^{\tilde
K}$.
%, and our
%$t_{IA}{}^B$ is denoted in CD as $\omega_{IA}{}^B$.
The object $\tilde B_{ab}$ is the one denoted as $B_{ab}$ in CD and in
GZ, while $\widehat{\mathcal{H}}_{ab}^{\widetilde{I}}$ is denoted as
$Q_{ab}^{\widetilde{I}}$ in CD. In CD there are two coupling constants,
$g$ and $g_R$, which we find to be the same.

Our results for vector and tensor multiplets without off-diagonal
vector-tensor couplings ($t_{IJ}{}^M=0$) agree with GZ. Adding the
couplings to hypermultiplets we also mostly agree with CD, though some
coefficients differ. For example, we differ in the coefficient in front
of the hyperino mass term and in the hyperino bilinear proportional to
the field strength ${\cal H}$.

However, the action~(\ref{final}) contains also more general couplings
than previously considered because of the possibility that
$t_{IJ}{}^M\neq 0$, i.e. a  representation of the gauge group on the
vector and tensor multiplets that is not completely reducible. For the
bosonic part of the action, this is explicitly visible in the
Chern-Simons couplings. Furthermore, the values of $K_I^x$ and of $W^x$
have implicit dependence on these representation matrices. The former
appear in the covariant derivatives of the scalars ${\mathcal D}_\mu \phi
^x$, thus directly influencing the fermionic supersymmetry
transformations, and the latter appear in the scalar potential.

%$h^I t_{IB}{}^A \bar{\zeta}_A \zeta^B$, is $-\sqrt{\ft32g\,{\rm i} $
% XXX $+\sqrt{6}g\,{\rm i} h^I t_{IB}{}^A \bar{\zeta}_A \zeta^B$\\
%
%\begin{table}[htb]
%\begin{minipage}{\linewidth}
%\renewcommand{\thefootnote}{\thempfootnote}
%\begin{center}
%\begin{tabular}{|c|c|}
%\hline
%this work&reference~\cite{Ceresole:2000jd}\\
%\hline\hline $-\ft3{8\sqrt{6}} \,{\rm i}
%h_{\widetilde{I}}H^{cd\widetilde{I}}\bar{\psi}^a \gamma_{abcd}\psi^b$&
%$+\ft3{8\sqrt{6}}\,{\rm i} h_{\widetilde{I}}H^{cd\widetilde{I}}\bar{\psi}^a \gamma_{abcd}\psi^b$\\
%$+\ft14\sqrt{\ft32}\,{\rm i} h_{\widetilde{I}}\bar{\zeta}_A\gamma \cdot
%H^{\widetilde{I}}\zeta^A$&
%$-\ft1{4\sqrt{6}}\,{\rm i} h_{\widetilde{I}}\bar{\zeta}_A\gamma \cdot H^{\widetilde{I}}\zeta^A$\\
%\hline
%\end{tabular}
%\caption{\it The table gives the translation between the conventions and
%objects used in this paper and the ones of
%\cite{Ceresole:2000jd}.\label{tbl:compareCD}}
%\end{center}
%\end{minipage}
%\end{table}
% The  gravitino bilinear we found
% agrees with the one of \cite{Gunaydin:1999zx}. The four-fermion terms
%were not compared. Finally, we mention that the first line in (4.10) in
%\cite{Ceresole:2000jd} has the opposite sign compared to ours.

%%%%%%%%%%%%%%%%%%%%%%%%%%%%%%%%%%%%%%%%%%%%%%%%%%%%%%%%%%%%%%%%%%%%%%%%%%%%%%%
\section{Simplified action for bosonic solutions}\label{s:simplified}
%%%%%%%%%%%%%%%%%%%%%%%%%%%%%%%%%%%%%%%%%%%%%%%%%%%%%%%%%%%%%%%%%%%%%%%%%%%%%%%
In applications where solutions to the five-dimensional $N=2$
supergravity theory are constructed, one is mainly concerned with the
bosonic terms in the action, as one often sets all fermions of the
solution equal to zero. A solution to the bosonic equations of motion
preserves $N$ supercharges if there are $N$ supersymmetry parameters
$\epsilon^i$ for which the corresponding supersymmetry variation of the
fermionic fields remains zero. Therefore, to search for supersymmetric
BPS solutions we only need to consider the bosonic action and the
supersymmetry rules of the fermions up to terms bilinear in the fermions.
For the convenience of the reader we collect these expressions in this
section. This section can be a starting point for a systematic search for
any kind of BPS solutions.

The bosonic part of the action reads
\begin{eqnarray}\label{truncated}
e^{-1}\mathcal{L}_{\rm bos}&=&
%Bosonic Kinetic terms
\ft1{2\kappa^2} R(\omega) -\ft14a_{\widetilde{I}\widetilde{J}}
H^{\widetilde{I}}_{\mu\nu}H^{\widetilde{J}\mu \nu} -\ft12
g_{xy}\mathcal{D}_\mu \phi^x\mathcal{D}^\mu \phi^y
-\ft1{2}g_{XY}\mathcal{D}_\mu q^{X}\mathcal{D}^\mu q^{Y}\nonumber\\&& +\,
\frac {1}{16 g} e^{-1} \varepsilon ^{\mu \nu \rho \sigma \tau}
\Omega_{MN} \widetilde{B}_{\mu \nu}^M \Big(\partial_\rho
\widetilde{B}_{\sigma \tau}^N+2g t_{IJ}{}^NA_\rho ^I F_{\sigma \tau}^J
   + g t_{IP}{}^N A_\rho^I \widetilde{B}_{\sigma\tau}^P\Big) \nonumber\\&&
%Potential
+\,\ft{g^2}{\kappa^4}\left( 4\vec{P}\cdot \vec{P}
-2\vec{P}^x\cdot\vec{P}_x- 2 W_x W^x
-2\mathcal{N}_{iA}\mathcal{N}^{iA}\right)\nonumber\\&&
%Chern-Simons terms
+\ft{\kappa}{12}\sqrt{\ft23} e^{-1} \varepsilon ^{\mu \nu \lambda \rho \sigma } {\cal C}_{IJK} A_\mu^I \left[ F_{\nu \lambda }^J F_{\rho \sigma }^K + f_{FG}{}^J A_\nu^F A_\lambda ^G \left(- \ft12 g  F_{\rho \sigma }^K + \ft1{10} g^2 f_{HL}{}^K A_\rho ^H A_\sigma^L \right)\right]\nonumber\\
& & - \ft{1}{8} e^{-1} \varepsilon^{\mu\nu\lambda\rho\sigma} \Omega_{MN}
t_{IK} {}^M t_{FG}{}^N A_{\mu}^I A_\nu^F A_\lambda^G \left(-\ft 12 g
F_{\rho\sigma}^K + \ft{1}{10} g^2 f_{HL}{}^K A_\rho^H A_\sigma^L \right).
\end{eqnarray}
The covariant derivatives in this bosonic truncation  are given by
\begin{equation}
\mathcal{D}_\mu \phi^x =\partial_\mu\phi^x+gA_\mu^IK_I^x  ,\qquad
\mathcal{D}_\mu q^{X} = \partial_\mu q^{X}+ g A_\mu^I k_I^{X}.
\end{equation}
The $N=2$ supersymmetry rules of the fermionic fields, up to bilinears in
the fermions, are given by
\begin{eqnarray}
\delta \psi_\mu^i&=&  D_\mu(\omega)\e^i + \ft{{\rm i}\kappa }{4\sqrt6}
h_{\widetilde I} H^{\widetilde I\nu \rho} \left( {\gamma}_{\mu\nu \rho }
- 4 g_{\mu \nu } {\gamma}_\rho\right) \e^i - \ft{1}{\kappa\sqrt6} {\rm i} g P^{ij}\gamma_\mu  \e_j ,\nonumber\\
\delta  \lambda^{xi} &=& - \ft{{\rm i}}{2} \slashed{\mathcal D} \phi^x
\e^i + \ft14 \gamma\cdot H^{\widetilde I}
 h^x_{\widetilde I} \e^i  - \ft{1}{\kappa^2} g P^{x\, ij} \e_j
 + \ft{1}{\kappa^2} g W^x \e^i ,\nonumber\\
\delta  \zeta^A&=& \ft12 {\rm i} \gamma^\mu \mathcal{D}_\mu q^X f_X^{iA}
\e_i + \ft{1}{\kappa} g {\mathcal N}_i^A \e^i  .
\end{eqnarray}
%%%%%%%%%%%%%%%%%%%%%%%%%%%%%%%%%%%%%%%%%%%%%%%%%%%%%%%%%%%%%%%%%%%%%%%%%%%%%%%
\section{Conclusions}\label{s:conclusions}
%%%%%%%%%%%%%%%%%%%%%%%%%%%%%%%%%%%%%%%%%%%%%%%%%%%%%%%%%%%%%%%%%%%%%%%%%%%%%%%
In this paper, we constructed  matter-coupled $N=2$ supergravity in five
dimensions, using the superconformal approach. For the matter sector we
took an arbitrary number of vector,  tensor and hypermultiplets. By
allowing off-diagonal vector-tensor couplings, we found more general
results than currently known in the literature. Our results provide the
appropriate starting point for a systematic search of BPS solutions such
as domain walls or black holes. Furthermore, they can be used to study
properties of the scalar potential that arises in flux-compactifications
of M-theory on Calabi-Yau manifolds, or any other five-dimensional
vacuum. The ingredients needed for such a search are given in the
previous section.

We end with some remarks.

For simplicity, we restricted ourselves to matter couplings for which an
action can be constructed. As pointed out in our previous paper
\cite{Bergshoeff:2002qk}, the superconformal approach also allows
conformal matter couplings for which no such action can be defined. These
theories are defined in terms of equations of motion that, without
introducing more variables, cannot be integrated to an action. It would
be interesting to extend the gauge-fixing procedure to these
theories\footnote{However, it would be natural to start also from
on-shell vector multiplets, as discussed in \cite{Gheerardyn:2003rf}.}.
In the present work we only introduced gauge-fixing conditions for
theories that have an action. In fact some of the gauge-fixing choices
were determined by requiring that the kinetic terms in the action had a
canonical (diagonal) form. It seems reasonable that precisely the same
conditions can also be applied to the theories without an action. This
would lead to new Poincar{\'e} matter couplings. It is known that there are
theories without an action that can be obtained via a Scherk-Schwarz
reduction of a theory with an action
\cite{Lavrinenko:1998qa,Bergshoeff:2002nv,Gheerardyn:2002wp,Hull:2003kr,Kerimo:2004qx}.
It would be interesting to see whether, in the same spirit, the $D=5$
matter couplings with no action can be obtained from some $D=6$ matter
couplings with an action.

Finally, a central role in this paper is played by the standard Weyl
multiplet. We showed that a second Weyl multiplet exists, the so-called
dilaton Weyl multiplet \cite{Bergshoeff:2002qk}. The two multiplets
describe the same number of degrees of freedom but differ in the field
content. {\sl A priori} the conformal programme can also be carried out
using the dilaton Weyl multiplet. It would be interesting to see whether
the dilaton Weyl multiplet may lead to matter couplings which cannot be
obtained by starting from the standard Weyl multiplet. In view of
previous results in four dimensions for $N=1$~\cite{Ferrara:1983dh} and
$N=2$~\cite{deWit:1985px} this is not expected, but it cannot be
excluded.

%%%%%%%%%%%%%%%%%%%%%%%%%%%%%%%%
\medskip
\section*{Acknowledgments}

\noindent We are grateful to Anna Ceresole, Gianguido Dall'Agata and Rein
Halbersma for useful discussions. This work is supported in part by the
European Community's Human Potential Programme under contract
HPRN-CT-2000-00131 Quantum Spacetime, in which E.B. and T.d.W. are
associated with the University of Utrecht. J.G. is Aspirant
FWO-Vlaanderen. The work of E.B. and T.d.W. is part of the research
program of the `Stichting voor Fundamenteel Onderzoek der materie' (FOM).
The work of S.C., J.G. and A.V.P. is supported in part by the Federal
Office for Scientific, Technical and Cultural Affairs through the
"Interuniversity Attraction Poles Programme -- Belgian Science Policy"
P5/27.
\newpage

%%%%%%%%%%%%%%%%%%%%%%%%%%%
\appendix

%%%%%%%%%%%%%%%%%%%%%%%%%%%%%%%%%%%%%%%%%%%%%%%%%%%%%%%%%%%%%%%%%%%%%%%%%%%%%%%
\section{Notation}
\label{app:notations}
%%%%%%%%%%%%%%%%%%%%%%%%%%%%%%%%%%%%%%%%%%%%%%%%%%%%%%%%%%%%%%%%%%%%%%%%%%%%%%%

\paragraph{Curvatures.}

First we explain our notation for the curvature tensors. We have Riemann
tensors defined on the five-dimensional spacetime and on the target space
manifold spanned by the hypermultiplet scalars. As we will show below, we
will use different conventions for the curvatures and Ricci tensors. They
have in common, however, that compact manifolds always have a positive
Ricci scalar curvature.

The `target space notation' starts from a connection denoted by
$\Gamma_{XY}{}^Z$, with Riemann curvature
\begin{equation}
 R_{XYZ}{}^W \equiv  2
\partial_{[X}\Gamma_{Y]Z}{}^W + 2 \Gamma_{V[X}{}^W \Gamma_{Y]Z}{}^V .
\label{defCurv}
\end{equation}
The Ricci scalar and Ricci tensor on the target space are then defined as
(in agreement with~\cite{Bergshoeff:2002qk})
\begin{equation}
  R= g^{XY}R_{XY}, \qquad R_{XY}= R_{ZXY}{}^Z.
 \label{Rtarget}
\end{equation}

Now we come to the definition of the spacetime curvature. In the general
relativity literature, one usually denotes the Levi-Civita connection as
$\Gamma^\rho{}_{\mu\nu}$, i.e. with the upper index on the left. The
Riemann curvature is defined as
\begin{equation}
  R^\sigma{}_{\rho\mu\nu}\equiv   2\partial_{[\mu} \Gamma^\sigma{}_{\nu]\rho
}+ 2 \Gamma^\sigma{}_{\tau [\mu }  \Gamma^\tau{}_{\nu ]\rho }  ,
 \label{Rspacetime}
\end{equation}
and has its upper index on the left, in contrast to (\ref{defCurv}).
Spacetime Ricci tensors and scalars are then defined as
\begin{equation}
  R= g^{\mu \nu }R_{\mu \nu }=g^{\mu \nu } R^\rho {}_{ \nu\rho  \mu}.
 \label{Riccispacetime}
\end{equation}
Note that this is a different convention from the one used
in~\cite{Bergshoeff:2001hc}, where the second and third indices are
contracted to define the Ricci tensor. This means that equations (3.11)
and (3.12) in~\cite{Bergshoeff:2001hc} change sign in the above
conventions. As stated before, we use these conventions such that compact
manifolds have a positive Ricci scalar curvature.

\paragraph{$\SU(2)$ and vector indices.}

At various places in the main text, we switch from SU(2) indices
$i,j=1,2$ to vector indices with the convention
\begin{equation}
A_i{}^j\equiv {\rm i} \vec A \cdot \vec \sigma_i{}^j,
\end{equation}
where $\vec \sigma$ are the Pauli matrices. With these conventions, we
obtain the identity
\begin{equation}
A_i{}^jB_j{}^k=-\vec A \cdot \vec B \delta_i{}^k -{\rm i} (\vec A \times
\vec B) \cdot {\vec \sigma}_i{}^k,
\end{equation}
for any two vectors $\vec A$ and $\vec B$.

Lowering and raising $\SU(2)$ indices is done using the $\varepsilon$
symbol, in northwest-southeast (NW-SE) conventions,
\begin{equation}
A^i=\varepsilon^{ij}A_j, \qquad A_i=A^j\varepsilon_{ji}.
\end{equation}
When the $\SU(2)$ indices are omitted (e.g. in spinor contractions),
NW-SE contractions are understood. For more details on the notation and
conventions about spinors, we refer to~\cite{Bergshoeff:2001hc}.

%%%%%%%%%%%%%%%%%%%%%%%%%%%%%%%%%%%%%%%%%%%%%%%%%%%%%%%%%%%%%%%%%%%%%%%%%%%%%%%
\section{Conformal multiplets}
\label{app:conf-matter}
%%%%%%%%%%%%%%%%%%%%%%%%%%%%%%%%%%%%%%%%%%%%%%%%%%%%%%%%%%%%%%%%%%%%%%%%%%%%%%%

In the following subsections we will repeat the relevant results for the
vector-tensor multiplet and hypermultiplet from~\cite{Bergshoeff:2002qk}.
Just like in~\cite{Bergshoeff:2002qk} we will discuss the two cases, with
or without metric, separately.

%%%%%%%%%%%%%%%%%%%%%%%%%%%%%%%%%%%%
\subsection{Weyl multiplet}
%%%%%%%%%%%%%%%%%%%%%%%%%%%%%%%%%%%%
The $Q$- and $S$-supersymmetry and $K$-transformation rules for the
independent fields of the standard Weyl multiplet
are~\cite{Bergshoeff:2001hc}
\begin{eqnarray}
\delta e_\mu{}^a   &=&  \ft 12\bar\e {\gamma}^a \psi_\mu , \nonumber\\
\delta \psi_\mu^i   &=& {\cal  D}_\mu \e^i + {\rm i} \gamma \cdot T
{\gamma}_\mu \e^i -{\rm i} {\gamma}_\mu \eta^i,\nonumber\\
\delta V_\mu{}^{ij} &=& -\ft32\,{\rm i} \bar\e^{(i} \phi_\mu^{j)} +4
\bar\e^{(i}{\gamma}_\mu\chi^{j)}
  + {\rm i} \bar\e^{(i} \gamma \cdot T \psi_\mu^{j)} + \ft32{\rm i}
\bar\eta^{(i}\psi_\mu^{j)} ,\nonumber\\
\delta T_{ab}     &=&  \ft12{\rm i} \bar\e {\gamma}_{ab} \chi -\ft
{3}{32}\,{\rm i} \bar\e\widehat R_{ab}(Q)    ,\nonumber\\
\delta \chi^i     &=&  \ft 14 \e^i D -\ft{1}{64} \gamma \cdot \widehat
R^{ij}(V) \e_j
                   + \ft18{\rm i} {\gamma}^{ab} \slashed{D} T_{ab} \e^i
                   - \ft18{\rm i} {\gamma}^a D^b T_{ab} \e^i \nonumber\\
               &&  -\ft 14 {\gamma}^{abcd} T_{ab}T_{cd} \e^i + \ft 16 T^2 \e^i
             +\ft 14 \gamma \cdot T \eta^i  ,\nonumber\\
\delta D         &=&  \bar\e \slashed{D} \chi - \ft {5}{3} {\rm i}
\bar\e\gamma \cdot T
\chi - {\rm i}  \bar\eta\chi ,\nonumber\\
\delta b_\mu       &=& \ft 12 {\rm i}\bar\e\phi_\mu -2 \bar\e{\gamma}_\mu
\chi + \ft12{\rm i} \bar\eta\psi_\mu +2\Lambda_{K\mu}.
\label{modifiedtransf}
\end{eqnarray}
The covariant derivatives are
\begin{eqnarray}
  {\cal
D}_\mu \e^i & = & \partial_\mu \e^i+\ft12 b_\mu\e^i +\ft14
\omega_\mu^{ab}
{\gamma}_{ab}\e^i - V^{ij}_\mu\epsilon_j, \nonumber\\
D_\mu T_{ab}&=& \left( \partial_\mu -b_\mu\right) T_{ab}-2\omega_{\mu
[a}{}^c T_{b]c}-\ft12{\rm i} \bar\psi_\mu  {\gamma}_{ab} \chi +\ft
{3}{32}{\rm i} \bar\psi_\mu \widehat R_{ab}(Q),   \nonumber\\
D_\mu \chi^i &=& \left( \partial_\mu -\ft32b_\mu +\ft14\omega_\mu
{}^{ab}\gamma_{ab}\right) \chi ^i-V_\mu^{ij}\chi_j\nonumber\\
&& -\,\ft14 \psi_\mu^i D +\ft{1}{64} \gamma \cdot \widehat R^{ij}(V)
\psi_{\mu j}      - \ft18{\rm i} {\gamma}^{ab} \slashed{D} T_{ab}
\psi_\mu^i
                   + \ft18{\rm i} {\gamma}^a D^b T_{ab} \psi_\mu^i\nonumber\\
               &&  +\,\ft14 {\gamma}^{abcd} T_{ab}T_{cd} \psi_\mu^i - \ft16 T^2 \psi_\mu^i
             -\ft14 \gamma \cdot T \phi_\mu^i . \label{covder2}
\end{eqnarray}
The covariant curvatures $\widehat{R}(Q)$ and $\widehat{R}(V)$ are
\begin{eqnarray}
  \widehat{R}_{\mu\nu}{}^i(Q) &=& R_{\mu\nu}{}^i(Q) + 2{\rm i} \gamma \cdot T
{\gamma}_{[\mu} \p_{\nu]}^i, \nonumber\\
R_{\mu\nu}{}^i(Q) & = & 2\partial_{[\mu}\psi^i_{\nu ]} +\ft12 \omega
_{[\mu}{}^{ab} {\gamma}_{ab}\psi^i_{\nu ]} +b_{[\mu}\psi^i_{\nu ]}
-2V_{[\mu}{}^{ij}\psi_{\nu ]j} +
2{\rm i}{\gamma}_a\phi ^i_{[\mu }e_{\nu ]}{}^a,\nonumber\\
\widehat{R}_{\mu\nu}{}^{ij}(V)&=& R_{\mu\nu}{}^{ij}(V) - 8
\bar{\psi}^{(i}_{[\mu} {\gamma}_{\nu]} \chi^{j)} - {\rm i}
\bar{\psi}^{(i}_{[\mu} \gamma \cdot
T \psi_{\nu]}^{j)} ,\nonumber\\
 R_{\mu\nu}{}^{ij}(V) & = & 2\partial_{[\mu}
V_{\nu]}{}^{ij} -2V_{[\mu}{}^{k( i} V_{\nu ]k}{}^{j)} -3{\rm i}{\bar\phi
}^{( i}_{[\mu}{\psi}^{j)}_{\nu ]} . \label{hatRQV}
\end{eqnarray}
The expressions for the dependent fields are given in (\ref{transfDepF}).

%%%%%%%%%%%%%%%%%%%%%%%%%%%%%%%%%%%%
\subsection{Vector-tensor multiplet}
%%%%%%%%%%%%%%%%%%%%%%%%%%%%%%%%%%%%

The supersymmetry rules for the vector-tensor multiplet coupled to the
five-dimensional standard Weyl multiplet are given by
\begin{eqnarray}
\delta A_\mu^I
&=& \ft12 \bar{\e} {\gamma}_\mu {\psi}^I -\ft12 {\rm i} \sigma^I \bar\e {\psi}_\mu , \nonumber\\
\delta B_{ab}^M
&=& - \bar \e {\gamma}_{[a} D_{b]} {\psi}^M- \ft12 {\rm i} \sigma^M \bar \epsilon \widehat{R}_{ab}(Q) + {\rm i} \bar \epsilon  {\gamma}_{[a} \gamma \cdot T {\gamma}_{b]} \psi ^M\nonumber\\
&& + {\rm i} g \bar{\e} {\gamma}_{ab} t_{({\widetilde J}{\widetilde K)}}{}^M \sigma^{\widetilde J} {\psi}^{\widetilde K} + {\rm i} \bar{\eta}{\gamma}_{ab}{\psi}^M, \nonumber\\
\delta Y^{ij \widetilde I}
&=& -\,\ft12 \bar{\e}^{(i} \slashed{D} {\psi}^{j) \widetilde I} +\ft12 \,{\rm i} \bar\e^{(i}\gamma  \cdot T {\psi}^{j) \widetilde I} - 4 {\rm i} \sigma^{\widetilde I} \bar\e^{(i} \chi^{j)}\nonumber\\
& & -\,\ft12 {\rm i} g \bar \e^{(i} \left(t_{[{\widetilde J} {\widetilde
K]}}{}^{\widetilde I} - 3 t_{({\widetilde J}{\widetilde K)}}
{}^{\widetilde I} \right) \sigma^{\widetilde J} \psi^{j) {\widetilde K}} + \ft12 {\rm i} \bar{\eta}^{(i} {\psi}^{j) \widetilde I} ,\nonumber\\
\delta {\psi}^{i \widetilde I} &=& - \ft14 \gamma  \cdot \widehat{{\cal
H}}{}^{\widetilde I} \e^i -\ft12{\rm i} \slashed{D} \sigma^{\widetilde I}
\e^i - Y^{ij \widetilde I} \e_j + \sigma^{\widetilde I} \gamma  \cdot T
\e^i +\ft12 g t_{({\widetilde J}
{\widetilde K)}}{}^{\widetilde I} \sigma^{\widetilde J} \sigma^{\widetilde K} \e^i + \sigma^{\widetilde I} \eta^i , \nonumber\\
\delta \sigma^{\widetilde I} &=& \ft12 {\rm i} \bar{\e}
{\psi}^{\widetilde I} . \label{tensorlocal}
\end{eqnarray}
The (superconformal) covariant derivatives are given by
\begin{eqnarray} \label{localderiv-tensor}
D_\mu\, \sigma^{\widetilde I} &=& {\cal D}_\mu \sigma^{\widetilde I} - \ft12{\rm i} \bar{\psi}_\mu \psi^{\widetilde I} ,\nonumber\\[2pt]
{\cal D}_\mu \sigma^{\widetilde I} &=& (\partial_\mu - b_\mu) \sigma^{\widetilde I} + g t_{J\widetilde K}{}^{\widetilde I} A_\mu^J \sigma^{\widetilde K} ,\nonumber\\[2pt]
D_\mu \psi^{i \widetilde I}
&=& {\cal D}_\mu \psi^{i \widetilde I} + \ft14 \gamma  \cdot \widehat{{\cal H}}^{\widetilde I} {\psi}_\mu^i + \ft12{\rm i} \slashed{D} \sigma^{\widetilde I} {\psi}_\mu^i + Y^{ij \widetilde I} {\psi}_{\mu j} - \sigma^{\widetilde I} {\gamma}\cdot T {\psi}_\mu^i  \nonumber\\
&& -\,\ft12 g t_{({\widetilde J}{\widetilde K)}}{}^{\widetilde I} \sigma^{\widetilde J} \sigma^{\widetilde K} \psi_\mu ^i - \sigma^{\widetilde I} \phi_\mu^i  ,\nonumber\\
{\cal D}_\mu \psi^{i \widetilde I} &=& \left(\partial_\mu - \ft32 b_\mu +
\ft14 {\gamma}_{ab} \hat{\omega}_\mu {}^{ab}\right) {\psi}^{i \widetilde
I} - V_\mu^{ij} {\psi}_j^{\widetilde I} + g t_{J\widetilde
K}{}^{\widetilde I} A_\mu^J \psi^{i \widetilde K} .
\end{eqnarray}
The covariant curvature $\widehat{{\cal H}}^{\widetilde I}_{\mu \nu }$
should be understood as having components $\left(\widehat F^I_{\mu \nu},
B^M_{\mu \nu }\right)$ with $\widehat F^I_{\mu \nu}$ given by
\begin{equation}
\widehat{F}_{\mu\nu}^I = 2 \partial_{[\mu} A_{\nu]}^I + g f_{JK}{}^I
A_\mu^J A_\nu^K - \bar{\psi}_{[\mu} {\gamma}_{\nu]} {\psi}^I + \ft12 {\rm
i} \sigma^I \bar{\psi}_{[\mu} {\psi}_{\nu]} .
\end{equation}
Finally, $\widehat{R}_{ab}(Q)$ is the supercovariant gravitino curvature
defined in equations (3.4) and (3.18) of \cite{Bergshoeff:2001hc}.

The vector-tensor multiplet can be realized in the absence of an action.
In that case, the tensor part is realized on-shell and the corresponding
equations of motion are given in~\cite[(4.4)]{Bergshoeff:2002qk}.

In the presence of a fully symmetric tensor $C_{{\widetilde
I}\,{\widetilde J}\,{\widetilde K}}$ the superconformal invariant action
can be written down and it takes the form {\allowdisplaybreaks
\begin{eqnarray}\label{conf-VTaction}
\lefteqn{e^{-1} {\cal L}_{\scriptsize{\hbox{Vector--Tensor}}}^{\rm conf}
= \biggl[ \left(- \ft14 {\cal\widehat H}_{\mu\nu}^{ {\widetilde I}} {\cal\widehat H}^{\mu\nu { {\widetilde J}}} - \ft12 \bar{\psi}^{ {\widetilde I}} \slashed{D} {\psi}^{ {\widetilde J}} + \ft 13 \sigma^{\widetilde I} \Box^c \sigma^{\widetilde J} + \ft 16 D_a \sigma^{ {\widetilde I}} D^a \sigma^{ {\widetilde J}} +  Y_{ij}^{{\widetilde I}} Y^{ij {{\widetilde J}}}  \right) \sigma^{ {\widetilde K}}}\nonumber\\
& & -\,\ft 43 \sigma^{\widetilde I} \sigma^{\widetilde J}
\sigma^{\widetilde K} \left(D + \ft{26}{3} T_{ab} T^{ab} \right)
+ 4 \sigma^{\widetilde I} \sigma^{\widetilde J} {\cal\widehat  H}_{ab}^{\widetilde K} T^{ab}\nonumber\\
& & -\,\ft 18 {\rm i} \bar{\psi}^{ {\widetilde I}} {\gamma}\cdot
{\cal\widehat H}^{ {\widetilde J}} {\psi}^{ {\widetilde K}} -\ft12 {\rm
i} \bar{\psi}^{i { {\widetilde I}}} {\psi}^{j { {\widetilde J}}}
Y_{ij}^{{\widetilde K}}  + {\rm i} \sigma^{\widetilde I}
\bar{\psi}^{\widetilde J}
{\gamma}\cdot T {\psi}^{\widetilde K} - 8 {\rm i} \sigma^{\widetilde I} \sigma^{\widetilde J} \bar{\psi}^{\widetilde K} \chi\nonumber\\
& & {+}\, \ft 16 \sigma^{\widetilde I} \bar {\psi}_\mu {\gamma}^\mu
\left({\rm i} \sigma^{\widetilde J} \slashed{D} {\psi}^{\widetilde K} +
\ft12 {\rm i} (\slashed{D} \sigma^{\widetilde J}) {\psi}^{\widetilde K} -
\ft 14 {\gamma}{\cdot} {\cal\widehat  H}^{\widetilde J} {\psi}^{\widetilde K}
+ 2 \sigma^{\widetilde J} {\gamma}{\cdot} T {\psi}^{\widetilde K} - 8 \sigma^{\widetilde J} \sigma^{\widetilde K} \chi \right) \nonumber\\
&&
{-} \ft 16 \bar {\psi}_a {\gamma}_b {\psi}^{\widetilde I} \left(\sigma^{\widetilde J} {\cal\widehat  H}^{ab {\widetilde K}} -8 \sigma^{\widetilde J} \sigma^{\widetilde K} T^{ab} \right) -\ft 1{12} \sigma^{\widetilde I} \bar {\psi}_\lambda  {\gamma}^{\mu\nu\lambda} {\psi}^{\widetilde J} {\cal\widehat  H}_{\mu\nu}^{\widetilde K} \nonumber\\
&& {+}\,\ft 1{12} {\rm i} \sigma^{\widetilde I} \bar {\psi}_a {\psi}_b
\left(\sigma^{\widetilde J} {\cal\widehat  H}^{ab {\widetilde K}} -8
\sigma^{\widetilde J} \sigma^{\widetilde K} T^{ab} \right) +\ft 1{48}
{\rm i} \sigma^{\widetilde I} \sigma^{\widetilde J} \bar {\psi}_\lambda
{\gamma}^{\mu \nu \lambda \rho } {\psi}_\rho {\cal\widehat
H}_{\mu\nu}^{\widetilde K}\nonumber\\&& {-}\, \ft12 \sigma^{\widetilde I}
\bar {\psi}_\mu^i {\gamma}^\mu {\psi}^{j \widetilde J} Y_{ij}^{\widetilde
K} +\ft 1{6} {\rm i} \sigma^{\widetilde I} \sigma^{\widetilde J}
\bar{\psi}_\mu^i {\gamma}^{\mu\nu} {\psi}_\nu^j Y_{ij}^{\widetilde K}
-\ft{1}{24} {\rm i} \bar {\psi}_\mu {\gamma}_\nu {\psi}^{\widetilde I}
\bar {\psi}^{\widetilde J} {\gamma}^{\mu\nu} {\psi}^{\widetilde
K}\nonumber\\&&  +\,\ft{1}{12} {\rm i} \bar {\psi}_\mu^i {\gamma}^\mu
{\psi}^{j {\widetilde I}} \bar {\psi}_i^{\widetilde J}
{\psi}_j^{\widetilde K} -\ft{1}{48} \sigma^{\widetilde I} \bar {\psi}_\mu
{\psi}_\nu \bar {\psi}^{\widetilde J} {\gamma}^{\mu\nu}
{\psi}^{\widetilde K} +\ft {1}{24} \sigma^{\widetilde I} \bar{\psi}_\mu^i
{\gamma}^{\mu\nu} {\psi}_\nu^j \bar{\psi}_i^{\widetilde J}
{\psi}_j^{\widetilde K} \nonumber\\&&
 -\,\ft 1{12} \sigma^{\widetilde I} \bar {\psi}_\lambda
{\gamma}^{\mu\nu\lambda } {\psi}^{\widetilde J} \bar{\psi}_{\mu}
{\gamma}_{\nu} {\psi}^{\widetilde K} + \ft 1{24}\,{\rm i}
\sigma^{\widetilde I} \sigma^{\widetilde J} \bar {\psi}_\lambda  {\gamma}^{\mu\nu\lambda} {\psi}^{\widetilde K} \bar{\psi}_{\mu} {\psi}_{\nu}\nonumber\\
& &
 + \,\ft 1{48} {\rm i} \sigma^{\widetilde I} \sigma^{\widetilde J} \bar {\psi}_\lambda  {\gamma}^{\mu\nu \lambda \rho } {\psi}_\rho \bar{\psi}_{\mu} {\gamma}_{\nu} {\psi}^{\widetilde K} + \ft1{96}\sigma^{\widetilde I} \sigma^{\widetilde J} \sigma^{\widetilde K} \bar {\psi}_\lambda  {\gamma}^{\mu\nu \lambda \rho } {\psi}_\rho  \bar{\psi}_{\mu} {\psi}_{\nu}\biggr]
  C_{{\widetilde I}{\widetilde J}{\widetilde K}}  \nonumber\\
 &&+ \,\ft {1}{16 g} e^{-1} \varepsilon ^{\mu \nu \rho \sigma \tau} \Omega_{MN} \widetilde{B}_{\mu \nu}^M
\left(
\partial_{\rho} \widetilde{B}_{\sigma \tau}^N+2g  t_{IJ}{}^NA_{\rho}^I F_{\sigma \tau}^J
   + g t_{IP}{}^N A_{\rho}^I \widetilde{B}_{\sigma \tau}^P\right)\nonumber\\
& &  -\,\ft 1{24} e^{-1} \varepsilon ^{\mu\nu \lambda \rho \sigma }
C_{IJK} A_\mu^I\!
 \left(\!F_{\nu\lambda}^J F_{\rho\sigma}^K - f_{FG}{}^J \!A_\nu^F A_\lambda ^G   \left(\ft12 g  F_{\rho \sigma }^K - \ft1{10} g^2 f_{HL}{}^K A_\rho ^H A_\sigma^L  \right)\right)
\nonumber\\
&& -\, \ft{1}{8} e^{-1} \varepsilon^{\mu\nu\lambda\rho\sigma} \Omega_{MN}
t_{IK}{}^M t_{FG}{}^N A_{\mu}^I A_\nu^F A_\lambda^G \left(-\ft12 g
F_{\rho\sigma}^K
+ \ft{1}{10} g^2 f_{HL}{}^K A_\rho^H A_\sigma^L \right)\nonumber\\
 &&+\, \ft14 {\rm i} g \bar{\psi}^{\widetilde I}\psi
^{\widetilde{J}}\sigma ^{\widetilde{K}}\sigma^{\widetilde{L}}\left(
t_{[\widetilde{I}\widetilde{J}]}{}^{\widetilde{M}}C_{\widetilde{M}\widetilde{K}\widetilde{L}}
-4t_{(\widetilde{I}\widetilde{K})}{}^{\widetilde{M}}C_{\widetilde{M}\widetilde{J}\widetilde{L}}
\right) \nonumber\\
&& -\, \ft14 g \bar{\psi}_\mu {\gamma}^\mu {\psi}^{\widetilde I}
\sigma^{\widetilde J} \sigma^{\widetilde K} \sigma^{\widetilde L}
t_{(\widetilde J \widetilde K)}{}^{\widetilde M}
C_{\widetilde M \widetilde I \widetilde L}\nonumber\\
&&  -\,\ft 12 g^2 \sigma^I \sigma^J \sigma^K \sigma^{\widetilde M}
\sigma^{\widetilde N} t_{J{\widetilde M}}{}^P t_{K{\widetilde N}}{}^Q
C_{IPQ} , \label{Lform3}
\end{eqnarray}
} where the superconformal d'Alembertian is defined as
\begin{eqnarray}
 \Box^c
\sigma^{\widetilde I}
&=& D^a D_a \sigma^{\widetilde I}\nonumber\\[2pt]
&=& \left( \partial^a  - 2 b^a + \omega_b^{~ba} \right) D_a
\sigma^{\widetilde I} + g t_{J \widetilde K}{}^{\widetilde I} A_a^J D^a
\sigma^{\widetilde K} - \ft{1}2{\rm i} \bar{\psi}_\mu D^\mu
{\psi}^{\widetilde I} - 2 \sigma^{\widetilde I} \bar{\psi}_\mu
{\gamma}^\mu \chi
\nonumber\\[2pt]
& & + \,\ft12 \bar{\psi}_\mu {\gamma}^\mu {\gamma}\cdot T
{\psi}^{\widetilde I} + \ft12 \bar{\phi }_\mu {\gamma}^\mu
{\psi}^{\widetilde I} + 2 f_\mu{}^\mu \sigma^{\widetilde I} -\ft12 g
\bar{\psi}_\mu {\gamma}^\mu t_{ J \widetilde K}{}^{\widetilde I}{\psi}^J
\sigma^{\widetilde K},
\end{eqnarray}
and we introduced a tensor field $\widetilde{B}^M_{ab}$ as
\begin{equation}
B_{ab}^M=\widetilde{B}_{ab}^M-\bar{\psi}_{[a}\gamma_{b]}\psi^M+\ft12 {\rm
i} \sigma^M\bar{\psi}_a\psi_b.
\end{equation}

%%%%%%%%%%%%%%%%%%%%%%%%%%%
\subsection{Hypermultiplet}
%%%%%%%%%%%%%%%%%%%%%%%%%%%

The supersymmetry rules for the hypermultiplet coupled to the
five-dimensional standard Weyl multiplet were found to be
\begin{eqnarray}
\delta q^X
&=& - {\rm i} \bar\epsilon^i \zeta^A f_{iA}^X,\nonumber\\
\widehat \delta \zeta^A &\equiv& \delta \zeta^A + \zeta^B\omega_{XB}{}^A
\delta q^X \nonumber\\
&=& \ft12 \,{\rm i} \slashed{D} q^X f_X^{iA} \epsilon_i
  - \ft13 {\gamma}\cdot T k^X f^A_{iX} \epsilon^i + \ft12 g \sigma^I k_I^X f_{iX}^A \epsilon^i + k^X f^A_{iX} \eta^i,
\end{eqnarray}
where $\widehat \delta \zeta^A$ is the covariant variation of $\zeta^A$,
see section~2.3.1 of \cite{Bergshoeff:2002qk}. The symmetries of the
system determine the (superconformal) covariant derivatives
\begin{eqnarray}
D_\mu q^X
&=& {\mathcal D}_\mu q^X + {\rm i} \bar{\psi}_\mu^i \zeta^A f_{iA}^X, \nonumber\\
{\mathcal D}_{\mu} q^X &=& \partial_\mu  q^X - b_\mu  k^X - V_\mu^{jk}
k_{jk}^X +
 g A_{\mu}^I k_I^X , \nonumber\\
D_\mu \zeta^A &=& {\mathcal D}_\mu \zeta^A - k^X f_{iX}^A \phi_\mu^i +
\ft12 \,{\rm i} \slashed{D} q^X f_{iX}^A \psi_\mu^i + \ft13 {\gamma}\cdot
T k^X f_{iX}^A \psi_\mu^i -  \ft12 g\sigma^I k_I^X f_{iX}^A \psi^i_{\mu} \nonumber\\
{\mathcal D}_\mu \zeta^A &=& \partial_\mu  \zeta^A + \partial_\mu  q^X
\omega_{XB} {}^A \zeta^B + \ft14 \hat{\omega}_\mu {}^{bc} {\gamma}_{bc}
\zeta^A - 2 b_\mu  \zeta^A + g A_\mu^I t_{IB}{}^A \zeta^B.
\end{eqnarray}

The equations of motion for $\zeta^A$ and $q^X$ were obtained by imposing
the superconformal algebra, and are given
in~\cite[(4.9-11)]{Bergshoeff:2002qk}.

Note that so far we did not require the presence of an action.
Introducing a metric, the locally conformal supersymmetric action is
given by
\begin{eqnarray}\label{conf-hyperaction}
e^{-1} \mathcal{L}_{\rm hyper}^{\rm conf} &=& - \ft12 g_{XY}
\mathcal{D}_a q^X {\mathcal{D}^a} q^Y +\bar{\zeta}_A \slashed{D} \zeta^A
+ \ft49 Dk^2 + \ft{8}{27} T^2 k^2
 \nonumber\\
&& -\, \ft{16}{3} {\rm i} \bar{\zeta}_A \chi^i k^X f_{iX}^A + 2 {\rm i}
\bar{\zeta}_A {\gamma}\cdot T \zeta^A
-\ft14 W_{ABCD} \bar{\zeta}^A \zeta^B \bar{\zeta}^C \zeta^D  \nonumber\\
&& -\, \ft29 \bar{\psi}_a {\gamma}^a \chi k^2 + \ft13 \bar{\zeta}_A
{\gamma}^a {\gamma}\cdot T \psi_a^i k^X f_{iX}^A
+ \ft12 {\rm i} \bar \zeta_A {\gamma}^a {\gamma}^b \psi_a^i \mathcal{D}_b q^X f_{iX}^A  \nonumber\\
&& + \, \ft23 f_a {}^a k^2 - \ft16 {\rm i} \bar{\psi}_a {\gamma}^{ab}
\phi_b k^2
 - \bar{\zeta}_A {\gamma}^a \phi_a^i k^X f_{iX}^A \nonumber\\
&& +\, \ft{1}{12} \bar{\psi}_a^i {\gamma}^{abc} \psi_b^j \mathcal{D}_c
q^Y J_Y {}^X {}_{ij} k_X - \ft19 \,{\rm i} \bar{\psi}^a \psi^b T_{ab} k^2
+ \ft{1}{18} \,{\rm i} \bar{\psi}_a {\gamma}^{abcd} \psi_b T_{cd} k^2
\nonumber\\
&& -\, g\biggl( {\rm i} \sigma^I t_{IB} {}^A  \bar{\zeta}_A \zeta^B + 2
{\rm i} k_I^X f_{iX}^A \bar{\zeta}_A \psi^{iI} +\ft12 \sigma^I k_I^X
f_{iX}^A \bar{\zeta}_A {\gamma}^a \psi_a^i
 \nonumber\\
&&\hphantom{- g\biggl(} + \bar{\psi}_a^i {\gamma}^a \psi^{jI}
P_{Iij}-\ft{1}{2} {\rm i} \bar{\psi}_a^i {\gamma}^{ab} \psi^j_b \sigma^I
P_{Iij}\biggr)
\nonumber\\
&&{}+2g Y^{ij}_I P_{ij}^I  -\ft12 g^2 \sigma^I \sigma^J k_I^X k_{JX}.
\end{eqnarray}

\section{Identities of very special geometry}
\label{app:veryspecial}

The vector multiplets are defined in terms of the symmetric real constant
tensor ${\cal C}_{IJK}$. The independent scalars are $\phi ^x$, but many
quantities are defined by functions $h^{I}(\phi)$, satisfying
\begin{equation}
 {\cal C}_{IJK} h^{I}(\phi )h^{J}(\phi)h^{K}(\phi )   =1,
\label{hhhC1}
\end{equation}
and
\begin{eqnarray}
  &&h_I(\phi )\equiv {\cal C}_{IJK}h^J(\phi )h^K(\phi)=a_{IJ}h^J,\qquad
  a_{IJ}\equiv -2{\cal C}_{IJK}h^K+3h_Ih_J,\nonumber\\
  && \Gamma_{IJK}\equiv
  -\sqrt{\ft23}\left({\cal C}_{IJK}-9h^L{\cal C}_{L(IJ}h_{K)}+9h_Ih_Jh_K\right), \quad
  R_{IJKL}=2\Gamma_{KM[J}\Gamma_{I]L}{}^M,
 \label{defquantI}
\end{eqnarray}
where here and below $I$-type indices are lowered or raised with $a_{IJ}$
or its inverse, which we assume to exist.

Define (with$_{,x}$ an ordinary derivative with respect to $\phi ^x$)
\begin{equation}
  h^I_x\equiv -\sqrt{\ft32}h^I_{,x}(\phi),
 \label{hIx}
\end{equation}
which, due to the constraint (\ref{hhhC1}) satisfies $h_Ih^I_x=0$,
leading to
\begin{equation}
  h_{Ix}\equiv a_{IJ}h^J_x=\sqrt{\ft32} h_{I,x}(\phi).
 \label{lowerhIx}
\end{equation}
We then also have
\begin{equation}
  h^I h_{Ix}=0, \qquad h_Ih^I{}_x=0.
 \label{hIhIx0}
\end{equation}
These quantities define the metric on the scalar space, which is the
pull-back of the metric $a_{IJ}$ to the subspace defined by
(\ref{hhhC1}):
\begin{equation}
  g_{xy}\equiv h_x^Ih_y^J a_{IJ}=-2h_x^Ih_y^J{\cal C}_{IJK}h^K.
 \label{gxy}
\end{equation}
The above relations can be written in matrix form
\begin{equation}
  \begin{pmatrix}h^I\cr h^I_x\end{pmatrix} a_{IJ}\begin{pmatrix}h^J&h^J_y\end{pmatrix}
  =\begin{pmatrix}1&0\cr 0&g_{xy}\end{pmatrix}.
 \label{matrixag}
\end{equation}
We can find the inverse of the first and third $(n+1)\times (n+1)$
matrices on the left-hand side (using $h_I^y\equiv g^{yx}h_{Ix}$)
\begin{equation}
  \begin{pmatrix}h^I\cr h^I_x\end{pmatrix}\begin{pmatrix}h_I&h_I^y\end{pmatrix}=\begin{pmatrix}1&0\cr 0&\delta
_x^y\end{pmatrix} \   \rightarrow\
\begin{pmatrix}h_I&h_I^x\end{pmatrix}\begin{pmatrix}h^J\cr h^J_x\end{pmatrix}=\delta
_I^J.
 \label{Inversehhx}
\end{equation}
Multiplying the latter equation with $a_{JK}$ leads to
\begin{equation}
  h_Ih_J+h_I^xh_{Jx}=a_{IJ}.
 \label{a=hh+hh}
\end{equation}
Using the decomposition of the unity as in (\ref{Inversehhx}), we can
write (with `;' a covariant derivative including a connection
$h_{Jx;y}=h_{Jx,y}-\Gamma_{xy}^zh_{Jz}$ such that $g_{xy;z}=0$)
\begin{eqnarray}
 h_{Ix;y} & = & \delta_I^Jh_{Jx;y}=\left(h_Ih^J+h_I^zh^J_z\right)h_{Jx;y}=\sqrt{\ft23}\left(
 h_Ih^J_yh_{Jx}+T_{xyz}h^z_I\right) =\sqrt{\ft23}\left( h_I g_{xy}+T_{xyz}h^z_I\right),  \nonumber\\
 h^I{}_{x;y} & = & -\sqrt{\ft23}\left( h^I g_{xy}+T_{xyz}h^{Iz}\right),  \nonumber\\
 T_{xyz} & \equiv  & \sqrt{\ft32}h_{Jx;y}h^J_z=-\sqrt{\ft32}h_{Jx}h^J_{z;y}=
 {\cal C}_{IJK}h^I_xh^J_yh_z^K,\nonumber\\
 &&\Rightarrow\quad \Gamma_{xy}^z=h^{Iz}h_{Ix,y}-\sqrt{\ft23}T_{xyw}g^{wz}=
 h_I^zh^I_{x,y}+\sqrt{\ft23}T_{xyw}g^{wz} .
 \label{Txyz}
\end{eqnarray}
The tensor $T_{xyz}$ is symmetric. Comparing (\ref{a=hh+hh}) and
(\ref{defquantI}), we obtain
\begin{equation}
  h_I^xh_{Jx}=-2{\cal C}_{IJK}h^K+2h_Ih_J,
 \label{hhChhh}
\end{equation}
whose covariant derivative with respect to $\phi ^y$ leads to
\begin{equation}
  T_{xyz}h_I^xh_J^z={\cal C}_{IJL}h^L_y+h_{(I}h_{J)y}.
 \label{Thh}
\end{equation}
Multiplying with another $h_K^y$, using again the two expressions for
$a_{IJ}$, leads to
\begin{equation}
  T_{xyz}h_I^xh_J^yh^z_K= {\cal C}_{IJK}+\ft32a_{(IJ}h_{K)}-\ft52h_Ih_Jh_K.
 \label{Thhh}
\end{equation}
The curvature is
\begin{equation}
  K_{xyzu}=R_{IJKL}h^I_xh^J_yh^K_zh^L_u=\ft43\left(
  g_{x[u}g_{z]y}+T_{x[u}{}^wT_{z]yw}\right).
 \label{Kxyzu}
\end{equation}

The domain of the variables should be limited to $h^I(\phi )\neq 0$ and
the metrics $a_{IJ}$ and $g_{xy}$ should be positive definite. Due to
relation (\ref{matrixag}) the latter two conditions are equivalent.

%%%%%%%%%%%%%%%%%%%%%%%%%%%%%%%%%%%%%%%%%%%%%%%%%%%%%%%
\providecommand{\href}[2]{#2}\begingroup\raggedright\endgroup

%\bibliography{refd5conf}
%\bibliographystyle{toine}
\end{document}